\documentclass[fleqn,usenatbib]{mnras}

\usepackage{amsmath}
\usepackage{mathptmx}
\usepackage{ulem}
\usepackage{txfonts}
\usepackage[T1]{fontenc}
\usepackage{ae,aecompl}
\usepackage[dvipsnames]{xcolor}
\usepackage{graphicx}
\usepackage[pdf]{pstricks}
\usepackage{epstopdf}
\usepackage{amssymb}
\usepackage{latexsym}
\usepackage{tikz-cd}
\usepackage{bm}
\usepackage{natbib,twoopt}
\newcommand{\be}{\begin{equation}}
\newcommand{\ee}{\end{equation}}
\newcommand{\dd}{{\rm d}}
\newcommand{\lp}{\left(}
\newcommand{\rp}{\right)}
\newcommand{\dts}{{\rm d}^{3}s}
\newcommand{\vs}{\textbf{s}}
\newcommand{\phzs}{photo-$z$s}
\newcommand{\Phzs}{Photo-$z$s}
\newcommand{\phz}{photo-$z$}
\newcommand{\szs}{spec-$z$s}
\newcommand{\sz}{spec-$z$}
\newcommand{\Szs}{Spec-$z$}
\newcommand{\hvr}{\hat{\textbf{$\Omega$}}}

\title[Cosmological information from the 2MPZ catalogue]{Extracting cosmological information from the angular power spectrum of the 2MASS Photometric Redshift catalogue}

\author[Balaguera-Antol\'{\i}nez et al.]{\parbox{\textwidth}{A. Balaguera-Antol\'{\i}nez$^{1,2}$\thanks{balaguera@iac.es}, M. Bilicki$^{3, 4, 5}$ \thanks{bilicki@strw.leidenuniv.nl},   E. Branchini$^{6,7,8}$\thanks{ebranchini@fis.uniroma3.it}, A. Postiglione$^{6}$\thanks{postiglione@fis.uniroma3.it}}\\
$^{1}$Instituto de Astrof\'{\i}sica de Canarias, s/n, E-38205, La Laguna, Tenerife, Spain\\
$^{2}$Departamento de Astrof\'{\i}sica, Universidad de La Laguna, E-38206, La Laguna, Tenerife, Spain\\
$^{3}$Leiden Observatory, Leiden University, Niels Bohrweg 2, NL-2333 CA Leiden, The Netherlands\\
$^{4}$National Centre for Nuclear Research, Astrophysics Division, P.O. Box 447, 90-950 \L{}\'{o}d\'{z}, Poland\\
$^{5}$Janusz Gil Institute of Astronomy, University of Zielona G\'ora, ul. Licealna 9, 65-417 Zielona G\'ora, Poland\\
$^{6}$Dipartamento di Fisica, Universit\'a degli Studi Roma Tre, Via della Vasca Navale 84, Rome 00146, Italy\\
$^{7}$INFN Sezione di Roma 3, Via della Vasca Navale 84, Rome 00146, Italy\\
$^{8}$INAF, Osservatorio Astronomico di Roma, Monte Porzio Catone, Italy}

\begin{document}
\label{firstpage}
\pagerange{\pageref{firstpage}--\pageref{lastpage}} 
\maketitle

\begin{abstract}
Using the almost all-sky 2MASS Photometric Redshift catalogue (2MPZ) we perform for the first time a tomographic analysis of galaxy angular clustering in the local Universe ($z<0.24$). We estimate the angular auto- and cross-power spectra of 2MPZ galaxies in three photometric redshift bins, and use dedicated mock catalogues to assess their errors. We measure a subset of cosmological parameters, having fixed the others at their Planck values, namely the baryon fraction $f_{b}=0.14^{+0.09}_{-0.06}$, the total matter density parameter $\Omega_{\rm m}=0.30\pm 0.06$, and the effective linear bias of 2MPZ galaxies $b_{\rm eff}$, which grows from $1.1^{+0.3}_{-0.4}$ at $\langle z \rangle=0.05$ up to $2.1^{+0.3}_{-0.5}$ at $\langle z\rangle=0.2$, largely because of the flux-limited nature of the dataset. The results obtained here for the local Universe agree with those derived with the same methodology at higher redshifts, and confirm the importance of the tomographic technique for next-generation photometric surveys such as Euclid or LSST.
\end{abstract}


\begin{keywords}
cosmology: - large-scale structure of Universe - observations - cosmological parameters, galaxies: photometry
\end{keywords}

\section{Introduction}

Cosmological probes like the baryonic acoustic oscillations \cite[BAO; e.g.][]{ehu,2005ApJ...633..560E, Cole2005, 2008MNRAS.390.1470S,2014MNRAS.441...24A} and redshift-space distortions \citep[RSD; e.g.][]{1987MNRAS.227....1K, 1998ApJ...498L...1S, 1998ASSL..231..185H,2008Natur.451..541G} can be used to simultaneously trace the expansion history of the Universe and the growth of cosmic structures. These probes, together with the measurements of the temperature fluctuations in the cosmic microwave background (CMB) \cite[e.g.][]{WMAP-9,Planck2015} and distance measurements to Supernovae Type Ia \cite[e.g.][]{Kowalski2008}, are exploited not only to constrain the fundamental cosmological parameters, but also to reveal the nature of dark energy and to tests the validity of General Relativity on cosmic scales \cite[e.g.][]{2014PhRvD..89d3509T, 2014MNRAS.443.1065B}.

BAOs and RSDs are inferred from the two and three--point statistics of mass tracers, both in configuration and in Fourier space \citep[see e.g.][]{1994MNRAS.267..785C, Percival2001, LahavSuto2004,2007ApJ...657..645P, 2017MNRAS.469.1738S}. So far, this has mainly been possible thanks to extensive observational campaigns such as the Sloan Digital Sky Survey \citep[SDSS,][]{York2000}, dedicated to measure angular positions and spectroscopic redshifts (\szs\ hereafter) of a large number of extragalactic objects over big cosmological volumes.

However, spectroscopic observations have their limitations in terms of sky coverage and number density of tracers for which redshifts can be measured in practice. Currently, the number of available \szs\ is about $3$ million, and this quantity is unlikely to grow by more than an order of magnitude in the coming years \citep{Peacock2016}. Photometric datasets, on the other hand, already include $\sim 10^9$ extragalactic sources, and this number is expected to increase dramatically in the next decade thanks to the ongoing and planned imaging surveys \citep{DES,Ivezic2008,laureijs,Pan-STARRS}. This difference stems from the comparatively longer observation time required to measure spectra, whereas sparse sampling is required to guarantee efficient selection of spectroscopic targets at moderate to large redshifts. As a result, outside of the local volume of $z<0.1$, \sz\ campaigns map only specific, colour-preselected sources, such as luminous red galaxies, emission line sources, or quasars \citep[e.g.][]{SDSS-IV}. This results in a low number density, limited completeness of tracers, and high shot-noise.

Another important difference between photometric and spectroscopic surveys is their typical sky coverage. The former are usually (much)  wider than the latter, since spectroscopic observations require a trade-off between area and depth. As a result, wide, almost full-sky, spectroscopic datasets like the 2MASS Redshift Survey \citep[2MRS,][]{huchra} or the IRAS PSCz \citep{PSCz} are much shallower and contain fewer objects than 
their full-sky photometric counterparts, such as the catalogues based on the 2-Micron All-Sky Survey \citep[2MASS,][]{skrutskie} or on the Wide-Field Infrared Survey Explorer \citep[WISE,][]{wright} measurements \citep[e.g.][]{Kovacs2015,bilicki16}.

While spectroscopic surveys remain the primary datasets for three dimensional (3D) clustering analyses, the availability of wide and deep photometric catalogues allows us to perform studies of 2D, i.e. angular, clustering over much larger volumes. Indeed, two-point angular correlation functions and angular power spectra (APS hereafter) were historically the first statistics used to investigate the properties of the large scale structure of the Universe \citep[e.g.][]{1973ApJ...185..413P,1973ApJ...185..757H,1974ApJS...28...19P,1977ApJ...212L.107D}. In particular, the APS is the natural tool to analyze full-sky catalogues since spherical harmonics constitute the natural orthonormal basis on the sphere. This consideration applies to wide spectroscopic samples too, in which case the Bessel functions are included to trace clustering along the radial direction. The so-called Fourier-Bessel decomposition \citep[][]{1994MNRAS.266..219F, 1995MNRAS.275..483H}, has been however seldom applied so far due to the computational cost of the technique \citep[e.g.][]{1999MNRAS.305..527T,2004MNRAS.353.1201P,2012A&A...540A..60L}.

The APS has been used to quantify the 2D clustering properties in many existing photometric catalogues \citep[e.g.][]{2004MNRAS.351..923B,2007MNRAS.374.1527B,2007MNRAS.378..852P,2011MNRAS.412.1669T,dePutter2012,Ho2012,Ho2015,2012ApJ...761...13S,2013MNRAS.428.3487H,Leistedt2013,Leistedt2014,NusserTiwari2015}. Although cosmological information can be extracted from purely 2D samples \citep[e.g.][]{2004MNRAS.351..923B,NusserTiwari2015}, much more stringent tests can be performed if some knowledge of clustering in the radial direction is also available. This is, in essence, the idea behind the tomographic approach, in which 2D clustering analyses are performed in different radial shells, both in terms of auto- as well as cross-correlations between the bins. The better the proxy for the radial distance, the thinner the shells, the closer to a full 3D study the tomographic analysis is \citep[e.g.][]{BlakeBridle2005,2012MNRAS.427.1891A,SalazarAlbornoz2014}. The tomographic approach to angular clustering is in particular possible thanks to the availability of photometric redshifts (\phzs\ ) estimated from multi-wavelength broadband photometry \citep{Koo1985}. Indeed, most of the tomographic clustering analyses have focused on the SDSS galaxy and quasar photometric catalogues, i.e. targeting objects at relatively large redshifts ($z>0.4$) and using much less than full-sky. The sky coverage aspect is rather crucial, since APS errors scale with the square root of the employed area \cite[e.g][]{1980lssu.book.....P, 2003moco.book.....D}. This is one of the reasons why surveys like Euclid \citep{laureijs} and the Large Synoptic Survey Telescope \citep[LSST,][]{LSST}, designed to map large portions of the sky at large depths, will adopt the tomographic analysis of APS as one of their main cosmological probes.

In the recent years, \phz\ catalogues covering the full extragalactic sky have become available \citep{bilicki14,bilicki16}. Although relatively local, as compared to for instance SDSS, these samples are much deeper than what is available from spectroscopic full-sky datasets such as 2MRS and PSCz, while giving access to much larger sky areas than SDSS or other ongoing photometric campaigns, such as DES. It is thus finally possible and timely to attempt a tomographic angular clustering analysis in the local Universe.

The general goal of this paper is to exploit a new, local \phz\ catalogue in order to advance our understanding of the low-redshift $(z<0.25)$ Universe through the analysis of its clustering properties. Previous analyses of the local Universe have either probed the 3D mass distribution over limited volumes using spectroscopic galaxy surveys such as QDOt \citep{1999MNRAS.308..897L}, PSCz, 2MRS, the 2dF Galaxy Redshift Survey \citep[2dFGRS,][]{colless}, the 6dF Galaxy Survey \citep[6dFGS,][]{jones09}, and SDSS, or the projected 2D distribution in photometric surveys such as the Automated Plate Measurement Galaxy survey \citep[APM,][]{1996MNRAS.283.1227M} or 2MASS. While waiting for the next generation of wide and deep spectroscopic surveys like Taipan galaxy survey \citep{2017PASA...34...47D} or the $4$-metre Multi-Object Spectroscopic Telescope \citep[4MOST,][]{2012SPIE.8446E..0TD}, that will allow us to investigate 3D clustering over large areas and out to relatively large redshifts, we aim at bridging the current gap between spectroscopic and photometric studies by performing a tomographic clustering analysis using the recently released 2MASS Photometric Redshift catalogue \citep[2MPZ,][]{bilicki14}. This dataset encompasses $\sim1$ million 2MASS sources within its completeness flux limit of $K\leq 13.9$ mag, and provides precise and accurate \phzs\ for all the sources. Our study can be seen as an extension of earlier tomographic analyses down to smaller redshifts and wider angular scales than based on SDSS material \citep[e.g.][]{2011MNRAS.412.1669T}, but it also adds tomography to 2D photometric studies which used low-redshift all-sky data without any $z$-binning \citep[e.g.][]{2005MNRAS.364..593F}.

\begin{figure*}
\includegraphics[width=18cm]{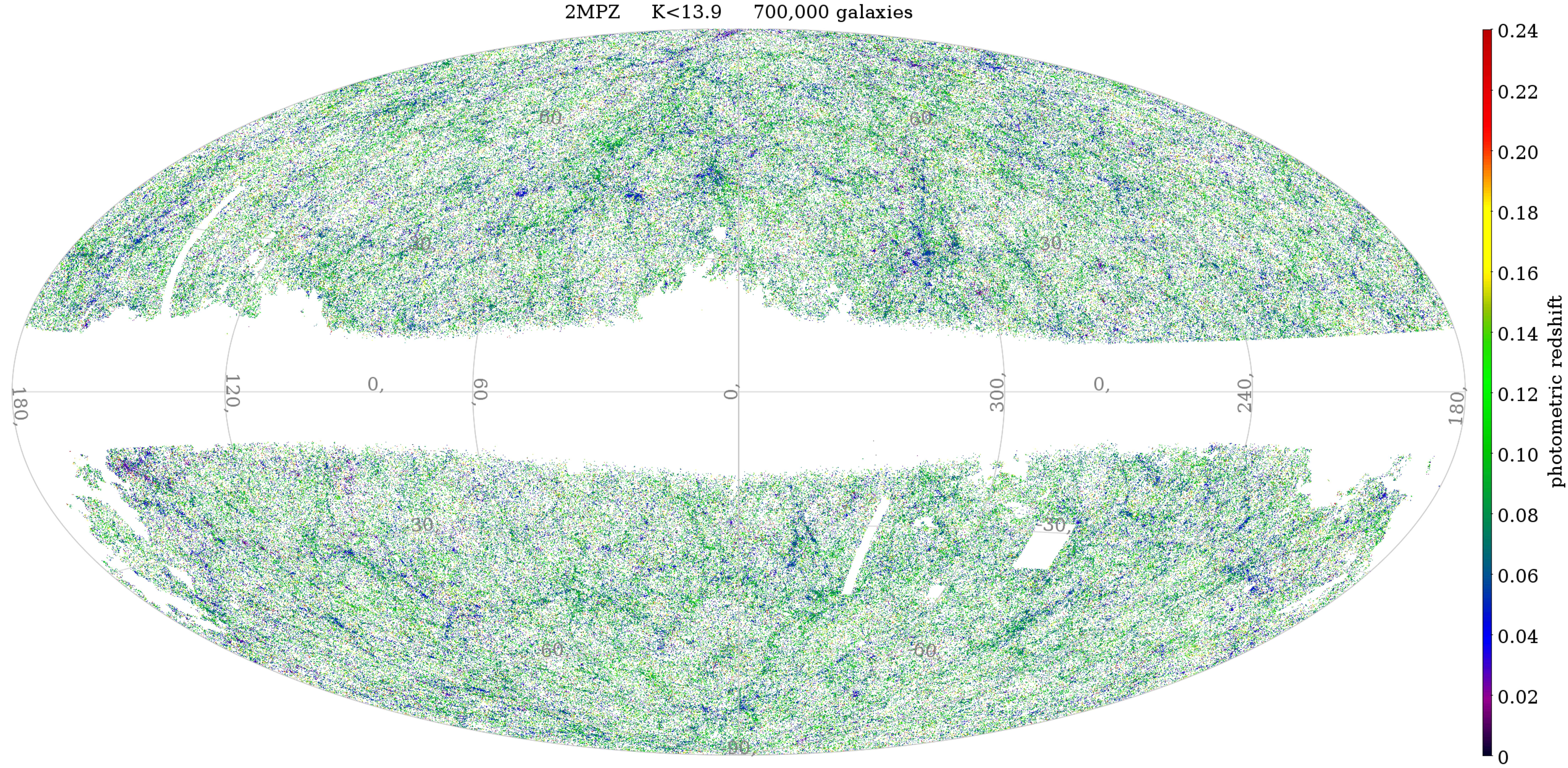}
\caption{Aitoff projection of the 2MPZ galaxy sample in Galactic coordinates. Colour coding in the bar identifies the \phz\ of the sources.}
\label{2mpz_gal}
\end{figure*}

The scientific motivations for performing this novel analysis are several. The most basic one is a quality check. A two-point clustering analysis is able to detect issues in a catalogue that evade other, more conventional investigations based on 1-point statistics, like number counts, luminosity functions as well as correlations among observed quantities, such as colour-colour or colour-magnitude diagrams. 2MPZ is a relatively new dataset in which \phzs\  have been measured using elaborate techniques potentially prone to systematic errors. Our analysis constitutes an additional and independent quality check for this catalogue.

A second goal closely related to the first one is to confirm or discard the presence of anomalies in the distribution of galaxies in the local Universe that have been hinted by previous analyses \citep[e.g.][]{2003MNRAS.345.1049F, 2005MNRAS.361..701F}. The most remarkable one is the alleged presence of an extended low density region in our cosmic neighborhood, the ``local hole'' \citep{2003MNRAS.345.1049F,2014MNRAS.437.2146W, 2016MNRAS.459..496W}, to which, however, our clustering analysis is not directly sensitive. Instead, we can focus on the second claimed anomaly, consisting of large power on wide angular scales, larger than expected in a $\Lambda$CDM Universe \citep[][]{2005MNRAS.361..701F}. Our tomographic analysis will be able to verify the reality of these earlier assertions better than what could be obtained from the original 2D analysis.

Our third and main goal is to obtain local estimates of cosmological parameters from a region that is significantly larger than those probed by spectroscopic surveys of low redshift objects. Matching results would constitute an important consistency check for the $\Lambda$CDM model. Similarly, and from a more methodological point of view, we shall compare our results with those of other tomographic analyses performed at larger redshifts \citep[e.g.][]{2007MNRAS.374.1527B, 2011MNRAS.412.1669T}. Because of this, we shall focus on the same, limited, subset of cosmological parameters that include the cosmological mean mass density, the baryon fraction, 
the {\it rms} density fluctuation of galaxy counts and the linear galaxy bias. The surveys considered in those analyses extended over smaller areas than our data but contained many more objects. We therefore expect the errors on our constraints to be larger and, for this reason, we decided not to extend our analysis to a larger set of cosmological parameters. 

Finally, we note that our analysis is somewhat complementary to the one recently performed by \cite{2018MNRAS.473.4318A} over the much shallower 2MRS sample (which however did not use the tomographic approach). While we focus on relatively large angular scales and the cosmological implications of the measured APS, the analysis of \cite{2018MNRAS.473.4318A} was aimed at characterising the typical environment of 2MRS galaxies through the same observable probed at smaller angular scales. Although in our analysis we can potentially characterize the 2MPZ environment in a similar way, we prefer to investigate the
issue in a follow-up paper in which we shall take advantage of the depth and number density of 2MPZ galaxies to push this type of analysis to larger redshifts and using different types galaxy populations within this sample.

  The outline of this paper is as follows. In Sect.~\ref{sec:cat} we describe the 2MPZ catalogue and the characterization of its photometric error distribution. That Section also presents the description of the mock catalogues used in the error analysis. In Sect.~\ref{sec:power} we briefly discuss the model of APS and the estimator implemented to analyze the 2MPZ catalogue. We present the measurements of APS in Sect.~\ref{results} and its covariance matrix. The Sect.~\ref{sec:lik} presents the likelihood analysis and constraints on cosmological parameters from the angular clustering of 2MPZ galaxies. We close with discussion and conclusions in Sect.~\ref{diss}.

Unless otherwise stated, throughout this work we adopt a fiducial, flat $\Lambda$CDM model with the same parameters as estimated by the Planck team \citep[][]{2014A&A...571A..16P}, namely, mean matter density $\Omega_{\rm m}=0.317$, baryon matter density $\Omega_{\rm b}=0.0489$, the amplitude of the primordial power spectrum at a pivot scale of $k=0.05\,h$ Mpc$^{-1}$, $10^{9}A_{s}=2.21$, the \textit{rms} of the matter distribution in spheres of $8$ Mpc $h^{-1}$ $\sigma_{8}=0.834$, the spectral index $n_{s}=0.963$, and the Hubble parameter $H_{0}=67.11$ km$/$s Mpc $h^{-1}$.


\section{The 2MASS Photometric Redshift catalogue}
\label{sec:cat}


\subsection{Description}\label{sec:description}
The 2MASS Photometric Redshift catalogue\footnote{Available for download from \url {http://ssa.roe.ac.uk/TWOMPZ.html}} \citep{bilicki14} is an almost all-sky flux-limited galaxy sample of $934,844$ objects in the \phz\ range $z_{\rm p} \in (0, 0.4)$ with $90\%$ of the sources within $z_{\rm p}<0.15$, and with mean redshift $\langle z_{\rm p} \rangle=0.07$. 2MPZ is the most comprehensive all-sky sample of the Universe in this redshift range to date. It can be regarded as an extension of the Two Micron All-Sky Survey \citep[2MASS,][]{skrutskie} Extended Source Catalogue \citep[XSC,][]{jarrett}. 

2MPZ was constructed by cross-matching 2MASS XSC with two additional all-sky data-sets, SuperCOSMOS XSC \citep{hambly,peacock} and WISE \citep{wright}. \Phzs\ have been estimated
for all the sources common to the three catalogues, using the ANNz \phz\ software \citep{annz}. Highly accurate \phz\ calibration was possible thanks to very comprehensive spectroscopic subsets of 2MASS, based on 2MRS, 6dFGS, 2dFGRS, and SDSS DR9 \citep{sdss}. They altogether encompass one-third of the whole 2MASS XSC and provide a very complete redshift training sample, especially thanks to SDSS. The resulting \phzs\ in 2MPZ are constrained to excellent precision and accuracy, with an overall mean bias of $\langle \delta z \rangle \sim 10^{-5}$ and random \phz\ error of $\sigma_{\delta z} \sim 0.013$ (see Sect.~\ref{sec:photoz} for a more comprehensive \phz\ error characterization). 2MPZ is flux-limited to $K\leq 13.9$ (Vega) which correspond roughly to the all-sky completeness limit of 2MASS XSC. Within this limit, 2MPZ includes $94\%$ of the 2MASS XSC objects. The missing sources are mostly located in areas not suitable for extragalactic science such as regions of high Galactic extinction, Magellanic Clouds, vicinity of bright stars, etc.

The incompleteness of 2MPZ with respect to 2MASS arises from the cross-match with the SuperCOSMOS and WISE datasets, which provide the multiband information needed to estimates \phzs. However, also the underlying 2MASS XSC is not complete all-sky, due to foreground contamination or confusion from our Galaxy or the Magellanic Clouds. In order to exclude regions with large incompleteness collectively called 'geometry mask', we proceeded as follows. We started by removing the areas in which either 2MPZ or 2MASS XSC are incomplete or contaminated, namely low Galactic latitudes ($|b|<10^\circ$), areas of high Galactic extinction ($EBV>0.3$ according to \citealt{SFD1998}) and of high stellar density ($\log n_{\mathrm star}\geq 3.5$, as derived from the 2MASS Point Source Catalogue\footnote{\url{https://www.ipac.caltech.edu/2mass/releases/allsky/doc/sec4_5c.html}}), as well as made manual cutouts of the Magellanic Clouds and stripes of missing WISE data due to `torque rod gashes'. We then used \texttt{Healpix} software \citep{2005ApJ...622..759G} to pixelate both 2MASS XSC and 2MPZ preselected in the same way at $K\leq 13.9$ and with all these above cutouts applied. By comparing number counts for each pixel we identified the sky areas which are incomplete in 2MPZ with respect to 2MASS. The resulting pixels were then added to the 2MPZ mask. This procedure automatically limits the maximum resolution of the mask, as to have enough statistics for the 2MASS vs.\ 2MPZ comparison, the \texttt{Healpix} $N_{\rm side}$ used was $64$ (pixel area of $\sim0.84$ deg$^2$), which was driven by the surface density of the two catalogues of $\sim 22$ sources per deg$^2$. See also \cite{alonso} for some more details; note however that the mask used there was slightly different than ours.

The $N_{\rm side}=64$ resolution of the mask gives $49152$ pixels, out of which $15104$ are  within the masked regions. The unmasked area corresponds to fraction $f_{\rm sky}\approx 0.69$ of the full sky, and contains $700,222$ galaxies up to $z_{\rm p}=0.24$, which represents the redshift of the most distant galaxy considered in our analysis. This redshift limit, together with the $K$-limit mentioned before, is what we define in this work as `the full sample'. In Fig.~\ref{2mpz_gal} we show the Aitoff projection in Galactic coordinates of the angular distribution of 2MPZ galaxies, colour-coded according to the \phz . The large scale features constituting the cosmic web are clearly seen despite projection effects \citep[see e.g.][for a description of the cosmic web as seen by 2MRS.]{2004PASA...21..396J}

\begin{figure}
  \includegraphics[width=8cm]{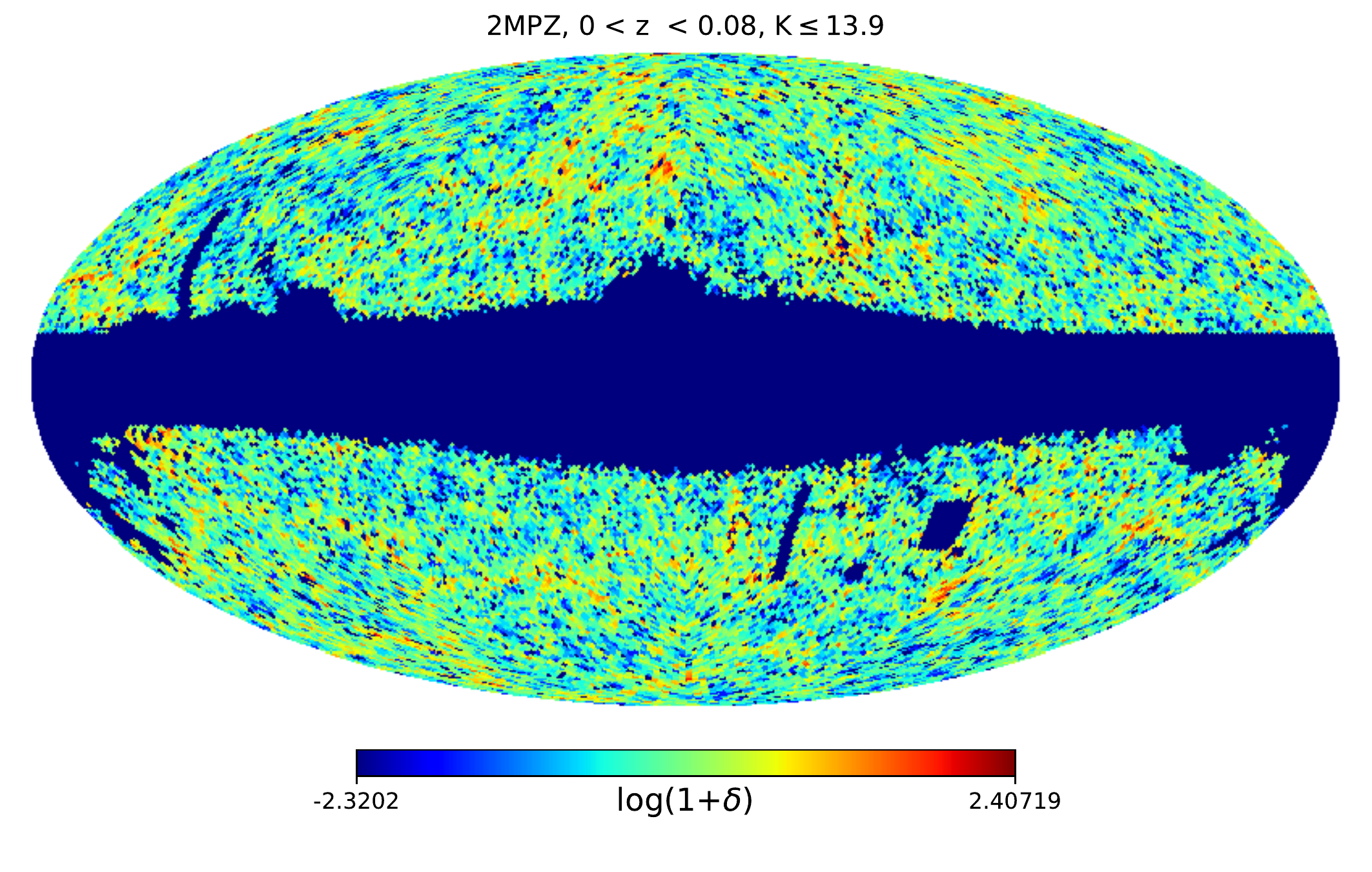}\\
  \includegraphics[width=8cm]{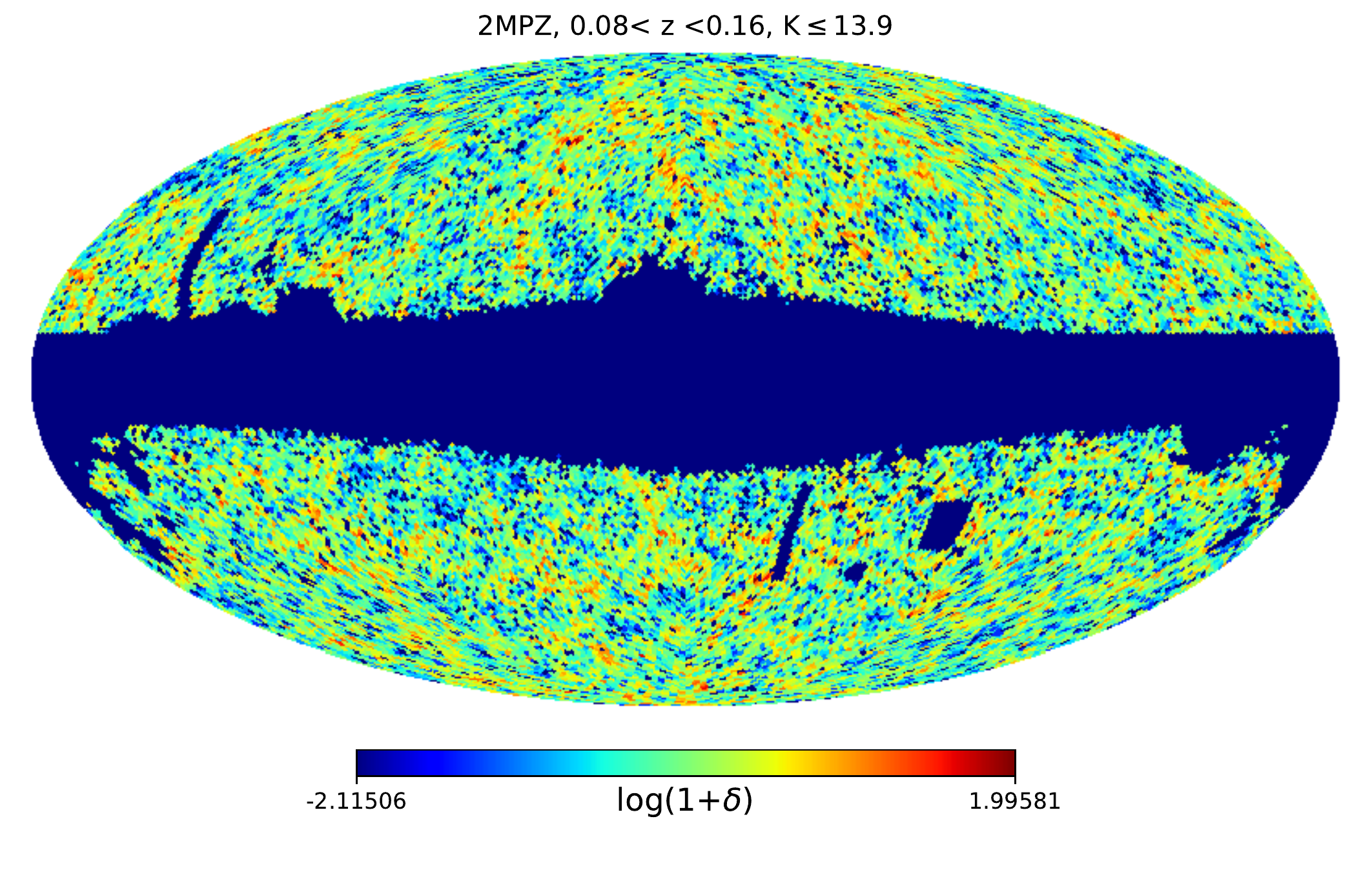}
 \includegraphics[width=8cm]{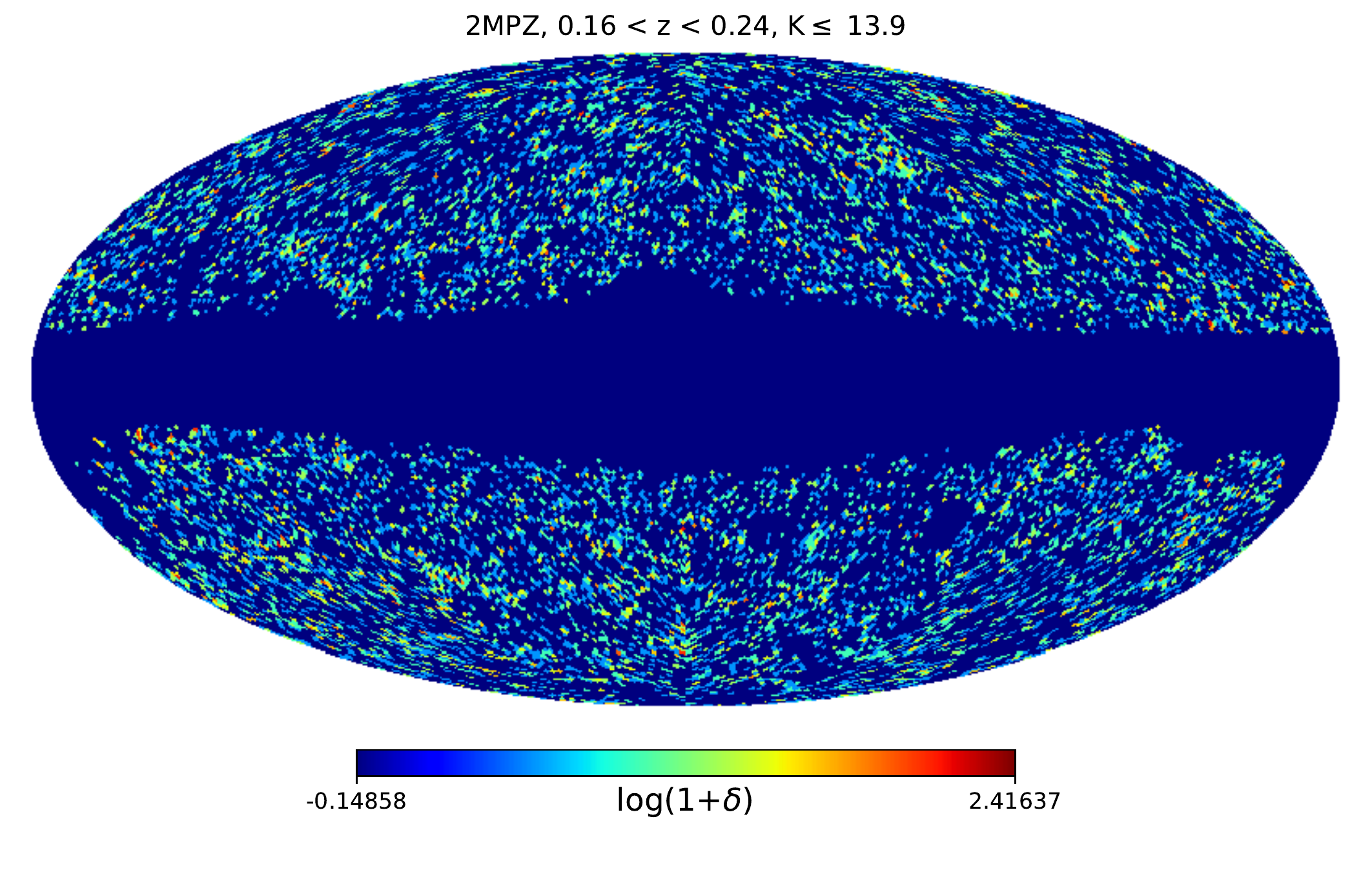}\\
 
 \caption{Mollweide projection in Galactic coordinates of the 2MPZ overdensity-map in three different \phz\ bins, indicated in the plots. The colour code shows the value of $\log(1+\delta_{i})$ in each pixel. 
 }\label{mask}
\end{figure}

It is worth stressing that the angular mask efficiently minimizes the impact of most systematic errors in the analysis of the angular clustering of 2MPZ galaxies, although it does not eliminate all of them. One example are coherent errors in the photometry, leading to a possibly varying depth of the dataset. In the 2MPZ case their main origin might be the fact that 2MASS and SuperCOSMOS input catalogues were both constructed by merging data from two telescopes observing two different hemispheres.

In the case of 2MASS, the two telescopes were identical \citep{skrutskie} and overlap among observations were large enough to guarantee a precise inter-calibration between hemispherical components. Nevertheless, due to different observational conditions at the two observational sites, the Northern (equatorial) part of the survey ($\delta > 12^\circ$) is deeper than the Southern one. This difference should be small at $K=13.9$, though not necessarily negligible.

\begin{figure}
  \vspace{-2cm}
\includegraphics[width=8.4cm]{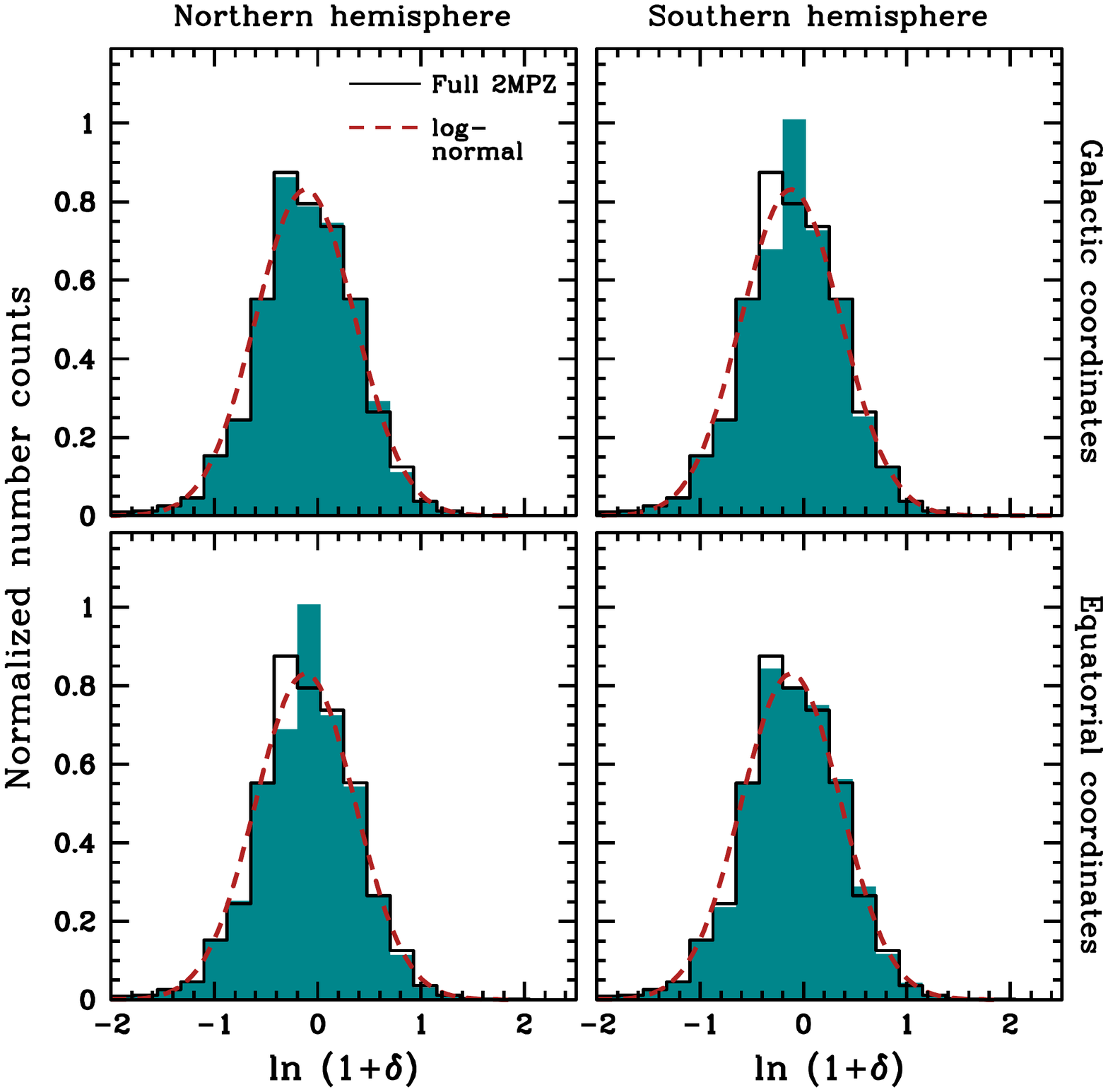}
\vspace{-2.5cm}
\caption{One-point PDF of the logarithmic density counts. Black solid-line histogram: full 2MPZ sample (the same in all the panels). Blue filled histograms: PDFs in different hemispherical subsamples identified by the labels in each panel. Red dashed curve: lognormal model with mean and variance 
computed from the full-sample counts (the same in all four panels).}\label{delta_dist}
\end{figure}

SuperCOSMOS is based on digitized scans of photographic plates from two hemispherical surveys, POSS-II and UKST, the split being at $\delta=2.5^\circ$. The two input samples were collected with different instruments, and colour-based calibration was essential to put the all-sky SuperCOSMOS magnitude measurements on a common scale. This calibration was fully completed only after the publication of the 2MPZ catalogue \citep{peacock}. What is more, after the 2MPZ sample had been published, it was recognized that the colour terms applied to SuperCOSMOS magnitudes in 2MPZ were partly incorrect \citep{bilicki16}, as were the extinction corrections in one of the hemispheres. These issues do not influence the sample selection itself (as it was based on 2MASS only), but can matter for the \phz\ estimation, which were calculated using eight photometric bands from 2MASS+WISE+SuperCOSMOS. We note however that the \phzs\ in 2MPZ were trained independently in the two hemispheres to self-calibrate such issues, so we expect them to be not significant. 
 
We believe that none of the systematics described above should be large enough to affect our clustering analysis. However, to guarantee that this is indeed the case, we have run a series of sanity checks in which we compared the APS measured in different sky areas (e.g. North vs. South hemispheres). The results of these tests are presented in Appendix \ref{check}. They confirm that no significant differences exist in the clustering properties of galaxies in different hemispheres. Although this does not rule out the presence of a large ``local hole'' \citep{2003MNRAS.345.1049F}, it certainly does not confirm its reality since one would expect that such a large underdensity would lead to significant variations of the galaxy clustering properties over very large scales.


\subsection{2MPZ galaxies: angular and redshift distribution}
\label{sec:mask_data}
In Fig. \ref{mask} we show \texttt{Healpix}-based Mollweide projections of 2MPZ galaxy surface overdensity, $\delta_{i}=N_{i}/\bar{N}-1$, where $N_{i}$ denotes the number of galaxies per pixel and $\bar{N}$ is the mean counts computed in three \phz\ intervals, indicated in the plots. Large scale features, corresponding to clusters and filaments, can be clearly identified, despite the thickness of the shell and projection effects. A simple visual inspection reveals therefore that a tomographic clustering analysis of 2MPZ galaxies should be indeed possible.

The width of redshift shells has been set equal to $\sim 5$ times the average \phz\ error. This choice represents a tradeoff between
the need to preserve clustering information along the line of sight (which requires narrow intervals) and that to minimize the contamination from objects in neighbouring redshift 
shells (which requires wide bins) \citep[][]{2011MNRAS.414..329C, 2011MNRAS.415.2193R}. 
In Table~\ref{tableres} we list the width of each redshift shell, the number of 2MPZ galaxies after masking, their surface density in the unmasked region, and the 
mean photometric galaxy redshift. The same quantities are also shown for the full 2MPZ sample (first row).
The last column lists the (Poisson) shot-noise correction that we apply to the APS estimated in each interval, as detailed in Sect.~\ref{sec:est}.

\begin{table}
\center
\begin{tabular}{|c|c|c|c|c|c|} \hline \hline
Redshift &    & $\langle z_{\rm p} \rangle $ &$N_{\rm gal}$  & $\bar{N}_{\rm gal}$& Shot \\
bins  &    &  &   & per deg$^{2}$ & noise \\ \hline
\blue{Full} &$(0,0.24)$ & $0.07$ &$700222$  & $24.8$ & $1.23\times 10^{-5}$ \\
\blue{z-bin} $1$&  $(0,0.08)$  & $0.056$ &$353530$  & $12.1$ & $2.53\times 10^{-5}$ \\
\blue{z-bin} $2$&  $(0.08,0.16)$ &  $0.109$ &$297318$  & $10.7$ & $2.83\times 10^{-5}$ \\
\blue{z-bin} $3$ &  $(0.16,0.24)$ &  $0.187$  &$49374$  & $1.7$ & $1.66\times 10^{-4}$\\
    \hline \hline
\end{tabular}\caption{\label{tableres} Catalogue statistics in the \phz\ bins considered in this analysis. The first row shows the full sample.}
\label{table}
\end{table}

The one-point probability distribution function (PDF hereafter) of the 2MPZ logarithmic surface density $\ln(1+\delta_{i})$ is shown in Fig.~\ref{delta_dist} 
(black solid line in all the panels) together with the best fit lognormal model (red dashed line) in which the mean and the variance are estimated from the counts. The PDF is approximately lognormal, which justifies the adoption of a lognormal PDF model in Sec.~\ref{smocks}.

In the same Figure, we compare the aforementioned PDF of the full sample with those from selected `hemispheres'. As is clear from the Figure, dividing the sample into two subsets (Northern vs. Southern hemisphere in both Galactic and Equatorial coordinates) does not affect significantly the PDF of the counts (blue filled histograms in the four panels), showing the same good match with the lognormal model as in the case of the full sample. This result indicates that systematic errors induced by photometric calibration issues are indeed small, as anticipated.

\subsection{2MPZ galaxies: redshift distribution and errors}
\label{sec:photoz}

\begin{figure}
    \vspace{-2cm}
  \includegraphics[width=9cm]{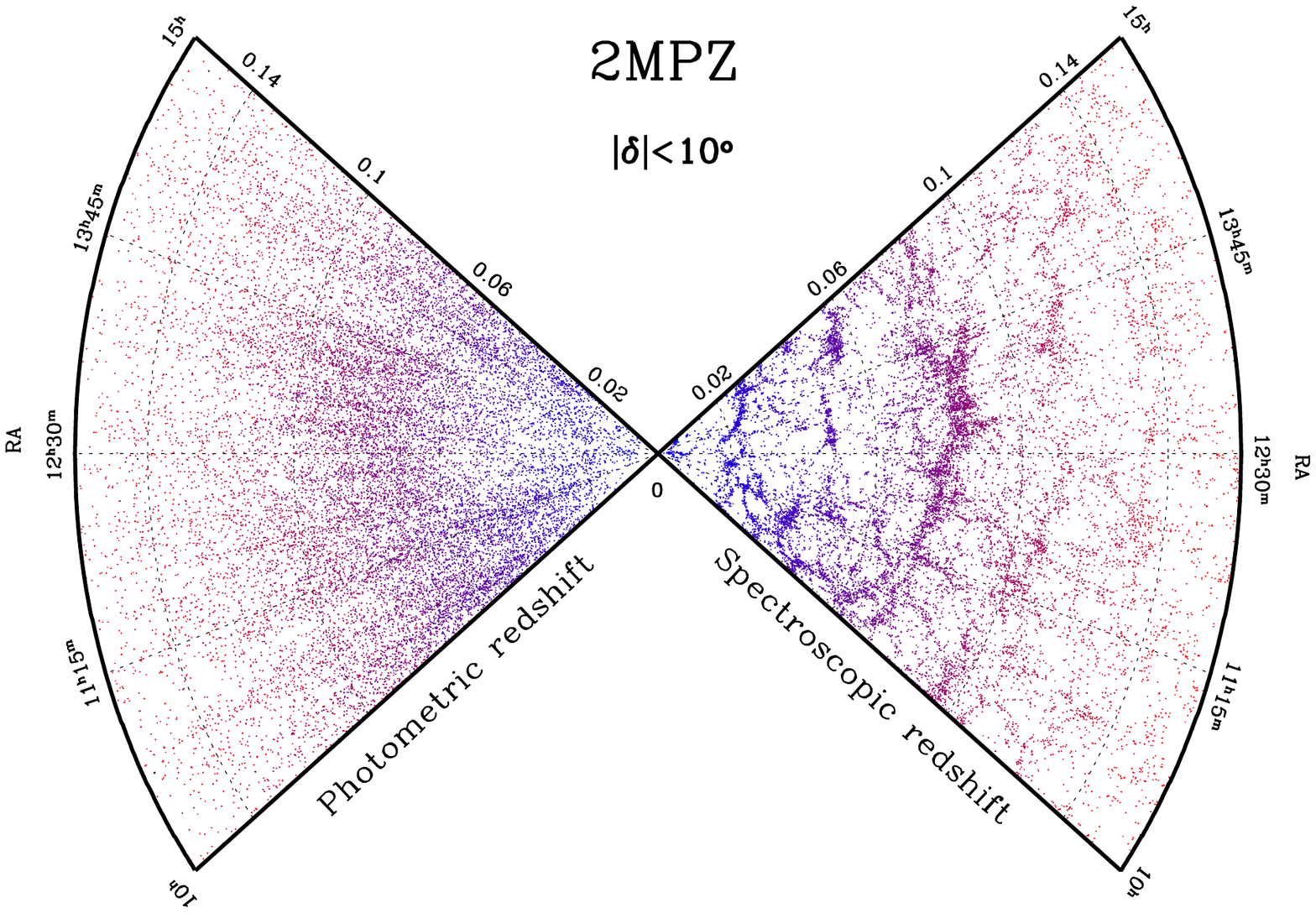}
  \vspace{-4.5cm}
  \caption{Pie diagram of a subsample of 2MPZ galaxies which have both spectroscopic and photometric redshift measured. Left: galaxy positions in \phz\ space.
  Right: galaxy positions in \sz\ space. The colour coding reflects \szs\ from light blue for nearby objects to dark red for distant galaxies.
  Colour mixing in the left panel further illustrates the effect of the {\it rms} random \phz\ error $\sigma_{z}\sim 0.01$.}
  \label{slice}
\end{figure}

Within the $K=13.9$ magnitude limit, $\sim 38\%$ of 2MPZ galaxies have both spectroscopic, $z_{\rm s}$, and photometric redshifts measured.
We use this overlap subsample to illustrate the effect of \phz\ errors on the measured clustering  in Fig.~\ref{slice}. The plot shows two ``pie diagrams"
representing the position of 2MPZ galaxies in a slice $|\delta| \leq 10^{\circ}$ thick in declination, and $75^{\circ}$ wide in 
right ascension. On the left hand side the radial position is assigned using the \phz\ as distance indicator. On the right hand side we use spectroscopic redshifts.
Errors on \phz\ obliterate the clustering signal on scales up to $50$ Mpc $h^{-1}$ along the line of sight, 
erasing prominent  structures such as the Sloan Great Wall \citep{2005ApJ...624..463G} at $z_{\rm s}\sim 0.08$. This observation qualitatively justifies 
the choice of \phz\ binning described in Sect.~\ref{sec:mask_data}.

\begin{figure}
  \vspace{-1cm}
    \includegraphics[width=8.9cm]{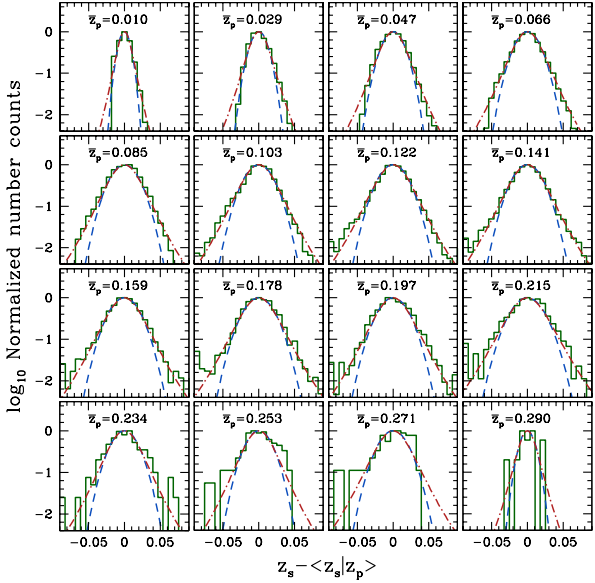}
    \vspace{-2.5cm}
  \caption{Distributions of the \phz\ errors, zPDF, as a function of  $z_{\rm s}-\langle z_{\rm s}|z_{p} \rangle$ in \phz\ bins of width $\Delta_{z}\sim 0.018$. The central redshift values of the bins, $\bar{z}_{p}$, 
   are indicated in the plot. Histograms: measured zPDF. Dashed curve: best fit Gaussian model with the same variance as the  measured zPDF. Dot-dashed curve: empirical zPDF model of Eq.~(\ref{varz}).}\label{pdfz}
\end{figure}

Because of the \phz\ errors, the observed redshift distribution of galaxies, $\dd N/\dd z_{\rm p}$, is different from the true one, $\dd N/\dd z_{\rm s}$. The relation between the two quantities is \citep[e.g.][]{2010MNRAS.403.2137S}:
\be\label{zdis_spec}
\lp\frac{\dd N}{\dd z_{\rm s}}\rp_{i}=\int _{0}^{\infty}W_{i}(z_{\rm p})\frac{\dd N}{\dd z_{\rm p}}P(z_{\rm s}|z_{\rm p})\,\dd z_{\rm p},
\ee
where $W_{i}(z_{\rm p})$ defines the photo-z bin, which in our case is a top-hat function. $P(z_{\rm s}|z_{\rm p})$ is the conditional probability (zPDF hereafter) of $z_{\rm s}$ given \ $z_{\rm p}$. To infer $\dd N/\dd z_{\rm s}$ (which is an input of our analysis) from the observed $\dd N/\dd z_{\rm p}$ we then need to estimate
zPDF. To do so, we consider the 2MPZ `overlap' subsample that have both $z_{\rm p}$ and $z_{\rm s}$. In order to highlight possible photo-z systematic errors, in Fig.~\ref{pdfz} we show, as green histograms, the zPDF as a function of $\delta z (z_{\rm p}) \equiv z_{\rm s}- \langle z_{\rm s}|z_{\rm p}\rangle$, where $\langle z_{\rm s}|z_{\rm p}\rangle$ is the mean \sz\ in a given bin of \phz\ . In each bin we measure the \textit{rms} scatter $\sigma^{2}_{z}(z_{\rm p})=\langle z_{\rm s}^2|z_{p}\rangle- \langle z_{\rm s}|z_{\rm p}\rangle^{2}$, which quantifies random errors. These are well fitted by $\sigma_{z}(z_{\rm p})\approx 0.03\tanh(-20.78 z_{p}^{2}+7.76z_{p} +0.05)$. They increase with the \phz\ from a value of $\sim 0.006$ at $z_{\rm p}\sim 0$ to $\sim 0.02$ at $z_{\rm p}\sim 0.24$. 

The dashed blue curves in Fig.~\ref{pdfz} represent Gaussian distributions with zero mean and a width $\sigma_{G}(z_{\rm p})\approx 0.9\sigma_{z}(z_{\rm p})/(1+z_{\rm p})$, which provides a good fit around the peak but fails to reproduce the extended tails of the distributions. Similarly as in \citet{bilicki14}, we also find that the function
\be\label{varz}
P(z_{\rm s}|z_{\rm p}) \propto \left[ 1+ \lp \frac{\delta z }{2\sigma_{G}(z_{\rm p})}\rp ^{2} \right]^{-3} \, ,
\ee
provides a better fit to the zPDF in all redshift bins, as is shown by the dot-dashed red curves in that Figure. 

\begin{figure}
\begin{center}
  \vspace{-2cm}
  \includegraphics[width=8.4cm]{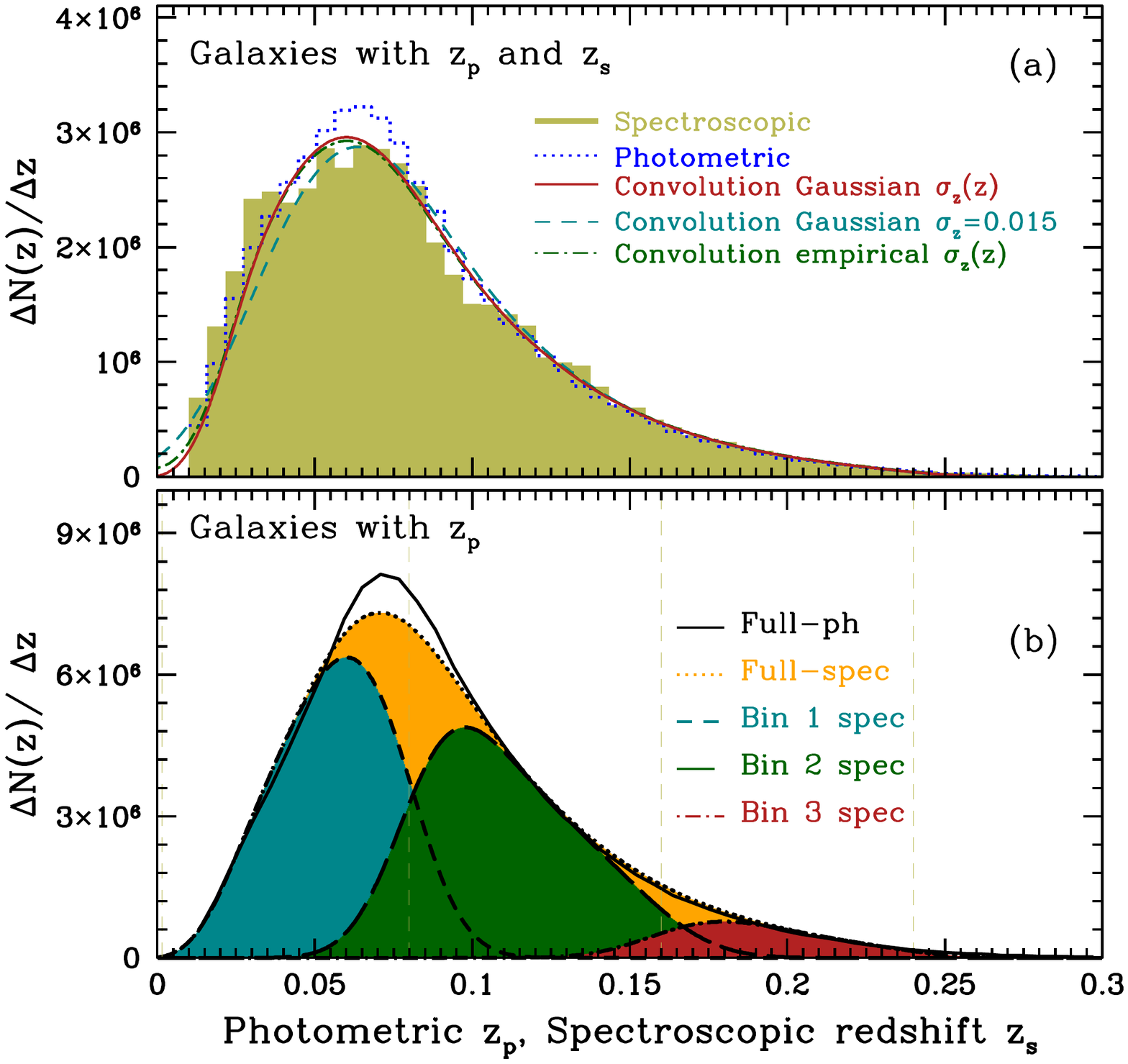}
\vspace{-2cm}
  \caption{Redshift distributions of 2MPZ galaxies. Top panel (a): 2MPZ galaxies in the overlap subsample with both spectroscopic, $z_{\rm s}$, and photometric redshifts, $z_{\rm p}$. Dotted, blue histogram:  $\dd N/\dd z_{\rm p}$.
    Filled, olive-green histogram: $\dd N/\dd z_{\rm s}$. Solid red, long-dashed blue and dot-dashed green curves: $ \dd N/\dd z_{\rm s}$ obtained assuming respectively a Gaussian error distribution zPDF with variable width (baseline), Gaussian with fixed  width, and the empirical model of Eq.~(\ref{varz}). Bottom panel (b): 2MPZ galaxies in the full sample. Black solid curve:  $\dd N/\dd z_{\rm p}$.
Orange dotted curve:  $\dd N/\dd z_{\rm s}$ inferred using the baseline zPDF. Other curves:  $\dd N/\dd z_{\rm s}$ of galaxies in the three photo-z bins
identified by the vertical dashed lines, obtained using the baseline zPDF.}\label{lf}
\end{center}
\end{figure}

The impact of \phz\ errors on the 2MPZ galaxy redshift distribution can be appreciated in Fig.~\ref{lf}. The top panel shows the $\dd N/\dd z_{\rm s}$ and $\dd N/\dd z_{\rm p}$ measured in the overlap subsample (filled and dotted histograms). The short-dashed curve illustrates the effect of convolving $\dd N/\dd z_{\rm p}$ with a Gaussian zPDF (Eq.~\ref{zdis_spec}) with fixed width equal to $0.015$. The inferred $\dd N/\dd z_{\rm s}$ underestimates the true one at small redshifts.
The continuous curve shows the effect of using a Gaussian zPDF with redshift-dependent width $\sigma_{G}(z_{p})$. The match with the observations improves considerably.

Using the zPDF from Eq.~(\ref{varz}) does not improve the quality of the fit further. As a consequence, we will model the zPDF as a Gaussian with redshift-dependent width. In doing this, we implicitly assume that the $\dd N/\dd z_{\rm s}$ of 2MPZ galaxies with both $z_{\rm p}$ and $z_{\rm s}$ measured is representative of the whole sample. This hypothesis is justified by the fact that a large part of the calibration data comes from SDSS, deeper and more complete than 2MPZ within their common area.

In the bottom panel of Fig. \ref{lf} we show the $\dd N/\dd z_{\rm p}$ of the full 2MPZ sample (black, continuous curve) and the inferred  $\dd N/\dd z_{\rm s}$ 
(dashed, orange curve), together with the $\dd N/\dd z_{\rm s}$ of the 2MPZ galaxies in the three photo-z bins identified by the vertical dashed lines.
As anticipated, the size of the bin guarantees an acceptable level of contamination from neighbouring redshift intervals.

\subsection{Mock 2MPZ galaxy catalogues}\label{smocks}
Previous analyses \cite[e.g.][]{2004MNRAS.351..923B, 2007MNRAS.374.1527B, 2011MNRAS.412.1669T} have assumed that errors on the APS are Gaussian. In this work we check the validity of this hypothesis 
by computing errors and their covariance from a suite of synthetic 2MPZ catalogues matching the properties of the real one.

Since a large number of independent mock catalogues are required to measure the covariance matrix with good accuracy\footnote{We are not aware of any existing $N$-body simulations which would allow us to select sufficiently many independent 2MPZ-like realizations for such an analysis.}, we shall make some assumptions
on the properties of these mocks. First of all, we assume that the mock galaxy density PDF is lognormal, which, as we have seen in Sect.~\ref{sec:cat}, is a good approximation. 
Furthermore, we assume that the $\ell$-modes of the mock 2MPZ angular spectrum measured over the full sky are all independent (i.e. we assume that mode-to-mode correlation is 
only induced by the geometry mask). Finally, as we are interested in measuring the angular spectrum in different redshift bins, we shall ignore any cross-correlation along the radial direction.

We generate the 2MPZ mock catalogues with the following procedure:
\begin{itemize}
\item We assume a fiducial cosmological model and compute the APS in the three redshift bins. We implement the public code \texttt{CLASSgal} \citep[][]{2013JCAP...11..044D}, which includes the nonlinear component of the dark matter power spectrum and corrections due to redshift space distortions (more details in Sect.~\ref{sec:power}).
\item We modulate the amplitude of the angular spectra to match the observed one (described in Sect.~\ref{sec:est}). With this procedure we implicitly determine the large-scale bias of the mock galaxies.
\item We generate Gaussian realizations of the angular spectrum in the three redshift bins and produce the corresponding \texttt{Healpix} surface density maps with a resolution matching that of the 2MPZ map described in Sect.~\ref{sec:mask_data}.
\item We perform a lognormal transformation which preserves the angular spectrum and obtain a lognormal PDF.
\item We impose the geometry of the 2MPZ sample represented by the mask described in Sec.~\ref{sec:description}.
\item We Monte-Carlo sample the maps to obtain a distribution of discrete objects in two steps: first, we assign \phz\ to an object according to the measured $\dd N/\dd z_{p}$; second, this object is assigned an angular position according to the angular surface density, which varies depending on the redshift bin in which the object is located. The number of mock objects in each redshift bin is drawn from a Poisson deviate with mean equal to the number of objects in the real sample.
\item \Szs\ are assigned following the results from Sec.~\ref{sec:photoz}.
\end{itemize}
We repeat the procedure until we generate $1000$ 2MPZ mock catalogues that we use to estimate errors in the measured angular spectrum and its covariance matrix.

Public codes such as \texttt{FLASK} \citep{2016MNRAS.459.3693X} can generate log-normal mock catalogues with correlation among different bins. In our likelihood analysis we verify that neglecting cross-correlation among \phzs\ in the 2MPZ clustering analysis does not affect significantly our results, thus justifying our choice for the construction of the mock catalogues.


\section{The angular power spectrum of 2MPZ galaxies}\label{sec:power}
In this Section we introduce the theory behind the model of the 2MPZ angular power spectrum and its estimator. The formalism and mathematical details can be found in, e.g. \citet[][]{1980lssu.book.....P, 1999coph.book.....P}.


\subsection{Modeling the angular power spectrum}\label{sec:power1}
The APS of galaxies with \sz\ in a given bin $i$ can be obtained from the harmonic decomposition of the observed surface density fluctuations around the mean $\bar{\sigma}_{i}$.
In case of a partial sky coverage, quantified by a binary angular mask $M(\hvr)$, the effective mean density depends on the direction: 
$\bar{\sigma}_{i}(\hvr)=\bar{\sigma}_{i}M(\hvr)$, where $\bar{\sigma}_{i}=N_{i}/\Delta \Omega$ is the mean surface density of $N_{i}$ over the 
unmasked area $\Delta \Omega$. The harmonic coefficients of the galaxy surface density fluctuation $\delta^{(i)}_{\rm gal}(\hvr)$ are
\be\label{almd}
a^{i,(s)}_{\ell m}=\int\, \delta^{(i)}_{\rm gal}(\hvr) Y^{*}_{\ell m}(\hvr) \,\dd \hvr = \int \dts\, \phi_{i}(\vs)\delta_{\rm gal}(\vs)Y^{*}_{\ell m}(\hvr),
\ee
where in the second expression the integral is in redshift space $\vs=z({\rm s},\hvr)$, $\phi_{i}(\vs)=\phi_{i}(s)M(\hvr)$ is the survey selection function in the
$i-$th redshift bin\footnote{The selection function is normalized in each bin such that $\int \phi_{i}(s) \, M(\hvr) \,  s^2 \, ds \, d\hvr =1$.} and $\delta_{\rm gal}(\vs)$ is the 3D galaxy density fluctuation. The first equality in this expression will be implemented to design the estimator of APS. The second one provides the starting point for the theoretical modeling of the APS.

Gravitational lensing, integrated Sachs Wolfe effect, and peculiar velocities modulate the observed galaxy density $\delta_{\rm gal}$. These effects need to be taken into account to obtain unbiased estimates of $a^{i,(s)}_{\ell m}$ \citep[e.g.][]{2011PhRvD..84d3516C}. At the low redshifts of the 2MPZ galaxies the dominant effect is peculiar velocities inducing RSD \citep[e.g.][]{1987MNRAS.227....1K, 1994MNRAS.266..219F,1995MNRAS.275..483H, 1996MNRAS.278...73H, 1998ASSL..231..185H}. We implement the public code \texttt{CLASSGal} \citep[][]{2013JCAP...11..044D}, in which the effect of the peculiar velocity field is computed from the cosmological parameters and no explicit parametrization of the RSD is done in terms of the linear redshift-space distortion parameter $\beta$ \citep[the ratio of the matter growth rate to the galaxy bias; e.g.][]{1987MNRAS.227....1K}. We use the options \texttt{`density'}, and/or \texttt{`rsd'} in order to account for real-space or redshift-space estimates of the angular power spectrum.

In general, the angular cross-spectrum between any two redshift bins $i$ and $j$ is:
\be
\label{cij2}
\tilde{C}^{ij}_{\ell }=\frac{1}{2\ell+1}\sum_{m=-\ell}^{\ell} \langle a^{i(s)}_{\ell m}a^{j (s) *}_{\ell m} \rangle=\sum_{\ell'}R_{\ell \ell'}C^{ij}_{\ell'}\,, 
\ee
where $R_{\ell \ell'}$ denotes the so-called \textit{mixing matrix}, which quantifies the effect of the geometry mask on the true power spectrum $C^{ij}_{\ell}$, the latter being expressed as
\be\label{ucl}
C^{ij}_{\ell}=b_{i}b_{j}\int_{0}^{\infty}\,\mathcal{P}(k)k^{2}F^{i}_{\ell}(k)F^{j}_{\ell}(k)\, \dd k\, .
\ee
In this expression $\mathcal{P}(k)$ is the three-dimensional, primordial matter power spectrum, $b_{i}$ is the linear bias of survey galaxies at $z=\langle z\rangle_{i}$.
The kernels $F^{i}_{\ell}(k)$ incorporates the effect of the survey selection function $\phi_{i}$, the matter transfer function $D(k,z)$ and RSD \cite[see e.g. equation $2.7$ of][]{ 2013JCAP...11..044D}. The version of these kernels written in terms of the parameter $\beta$ can be found, e.g. in equation $28$ of \citet[][]{2007MNRAS.378..852P}.


\subsection{2MPZ angular mixing matrix}\label{sec:mix}
\begin{figure}
\vspace{-2cm}
  \includegraphics[width=8.9cm]{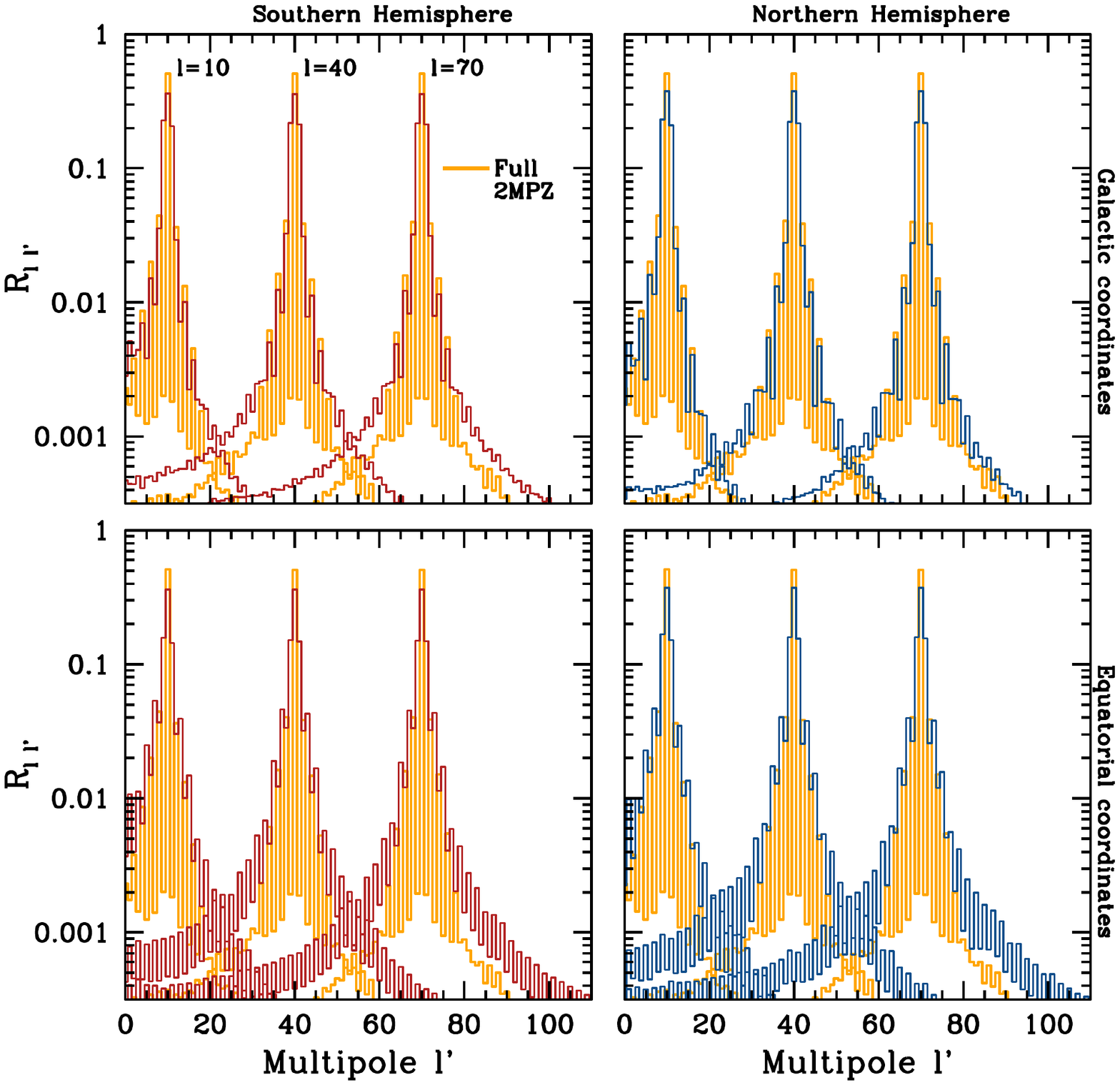}
\vspace{-2.5cm}
  \caption{Selected elements of the mixing matrix, $R_{\ell \ell'}$, computed using Eq.~(\ref{mm}), for the full 2MPZ survey (light histogram in all panels) and for the north and south hemisphere fractions in Galactic (top) and 
  Equatorial coordinates (bottom), as indicated in the plots (dark histograms).}\label{rll}
\end{figure}

The mixing matrix in Eq.~(\ref{cij2}) can be expressed in terms of the $3j$-Wigner symbols:
\be\label{mm}
R_{\ell \ell'}= \frac{(2\ell'+1)}{4\pi}\sum_{\ell''}(2\ell''+1)W_{\ell''}
\lp
\begin{array}{ccc}
\ell &\ell'&\ell''\\
0 & 0& 0\\
\end{array}
\rp ^{2}, 
\ee
where $W_{\ell}$ represents the APS of the geometry mask. In Fig.\ref{rll} we show some elements of the $R_{\ell \ell'}$ for the full 2MPZ mask (the light-coloured histogram in all panels) as well as those that refer to various half-sky samples (dark-coloured histograms in the different panels). The values of $\ell$ and $\ell'$ are indicated in the panels. Departures from $\delta-$Dirac shape indicate power leakage
from $\ell$ to $\ell' \ne \ell$. 
For the full 2MPZ case, and for the multipoles used in our analysis, $ \sim 75\%$ of power is preserved at the scale $\ell$ and $\sim 90\%$ is preserved in the range $\ell \pm 6$.
When only Northern and Southern hemispheres are used, the power preserved at the same multipole drops to $\sim 37\%$ in Galactic coordinates (upper panels) and to $\sim 35\%$ 
in Equatorial coordinates (bottom panels). This comparison highlights the importance of using an all-sky survey for such an analysis.
The precise figures are listed in Table \ref{rlld} together with the fraction of the unmasked sky, $f_{\rm sky}$, and the number of objects that it contains, $N_{\rm gal}$.

\begin{table}
\center
\begin{tabular}{|c|c|c|c|} \hline \hline
  \textbf{Hemisphere}  & $N_{\rm gal}$  & $f_{\rm sky}$ & \textbf{Fraction of} \\
   &   &  & \textbf{power at $\ell$} \\
  \blue{Full 2MPZ}  & $700222$ & $0.69$ & $75\%$ \\
  \blue{Northern Galactic}  & $360972$  & $0.35$ & $38\%$ \\
  \blue{Southern Galactic}  & $339250$  & $0.34$ & $36\%$ \\
  \blue{Northern Equatorial}  & $359507$  & $0.35$ & $37\%$ \\
  \blue{Southern Equatorial} & $340715$  & $0.34$ & $36\%$ \\
  \hline \hline
\end{tabular}\caption{Some characteristic of the 2MPZ angular mixing matrix, for hemispherical divisions in two coordinate systems, for the full \phz\ range.}
\label{rlld}
\end{table}

\subsection{Limber approximation and redshift space distortions}\label{rsddd}

\begin{figure}
  \vspace{-2cm}
  \includegraphics[width=8.7cm]{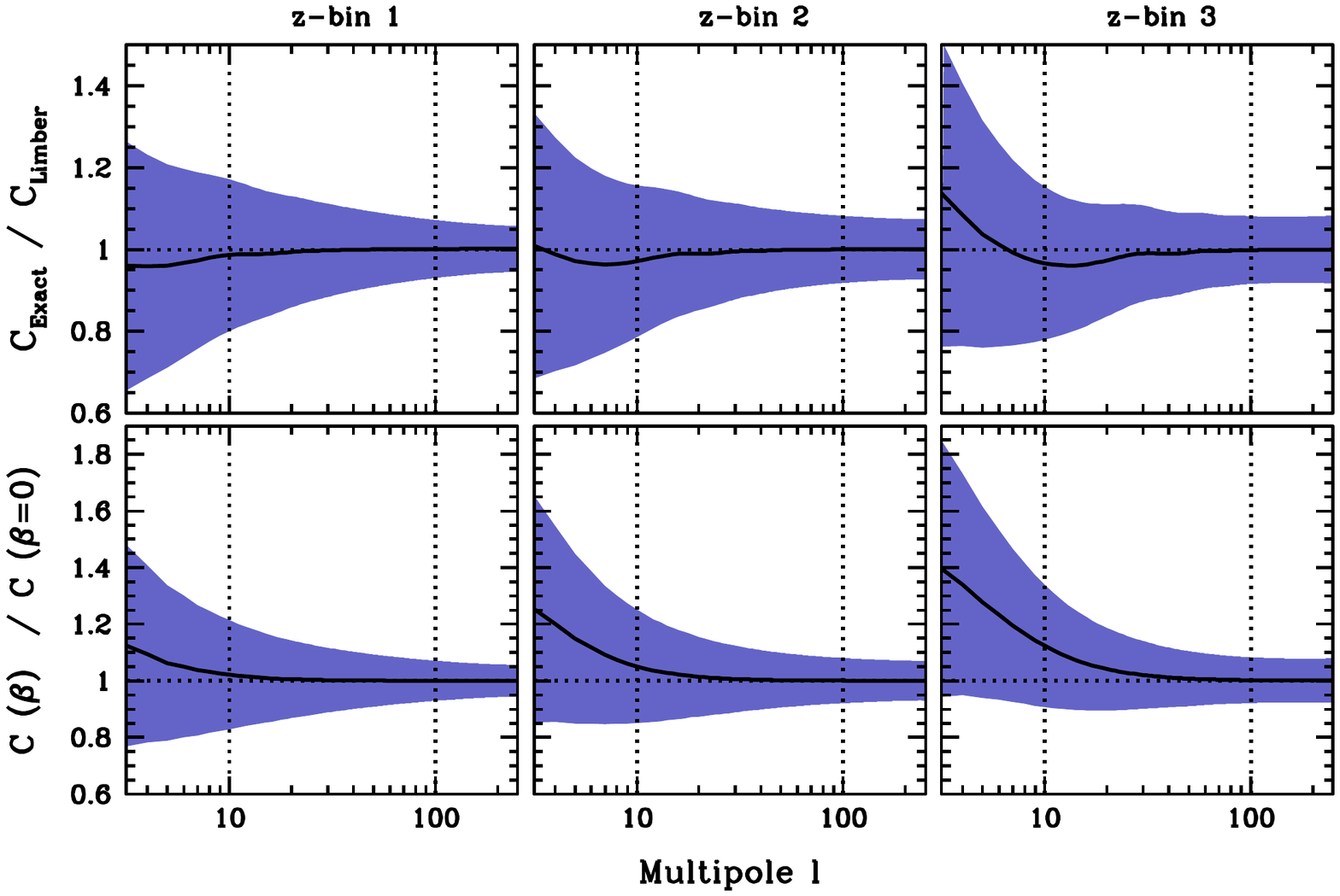}
  \vspace{-5cm}
  \caption{Top panels. Solid curve: bias introduced by the Limber approximation quantified by the ratio between the 
  exact 2MPZ angular spectrum of Eq.~(\ref{ucl}) and that obtained from Eq.~(\ref{limber}). 
  Bottom panels. Solid curve: RSD signature in the angular power spectrum  from the ratio between the 
  redshift and the real space angular spectra of 2MPZ galaxies. Shaded areas: Gaussian random errors.
  Panels from left to right indicate different redshift bins (see Table \ref{table}). All spectra have been computed using the same fiducial cosmological model convolved with the 2MPZ mixing matrix.}
  \label{class}
\end{figure}

The implementation of Eq.~(\ref{ucl}) involves the evaluation of spherical Bessel functions, which are computationally demanding. This is a potentially serious issue, since Eq.~(\ref{ucl}) needs to be evaluated for many different cosmological models when comparing observations with theory. Several methods have recently been proposed to mitigate this problem
 \citep[e.g.][]{2017arXiv170103592C, 2017arXiv170505022A}. Perhaps the most common approach is that of adopting the so-called Limber approximation \cite[e.g.][]{1953ApJ...117..134L, 2008PhRvD..78l3506L}, valid for $\ell\gg 1$. In this approximation Eq.~(\ref{ucl}) can be shown to reduce to
\be\label{limber}
C^{ij}_{\ell}\approx \frac{b_{i}b_{j}}{N_{i}N_{j}}\int_{0}^{\infty}\frac{\dd N_{i}}{\dd z}\frac{\dd N_{j}}{\dd z}P_{\rm mat}\lp\frac{\ell}{r(z)},z \rp \frac{H(z)}{r^{2}(z)}\,  \dd z,
\ee
where $H(z)$ is the Hubble function, $N_{i}=\int \dd z\, \dd N_{i}/\dd z$ is the expected number of galaxies in the $i-$th redshift bin and $P_{\rm mat}(k,z)=\mathcal{P}(k)D^{2}(k,z)$ is the matter power spectrum. The accuracy of this approximation depends on the angular scale, the cosmological model and the characteristics of the target galaxy sample such as the depth of the redshift shell and selection effects. 
The impact of using the Limber approximation for our study is shown in the top panels of Fig. \ref{class}, in which we plot the ratio of the exact expression for the angular spectrum for 2MPZ galaxies (Eq.~\ref{ucl}) and the one evaluated with Eq.~(\ref{limber}), in the three redshift bins considered in our analysis, for the fiducial cosmological model. Both spectra have been convolved with the same mixing matrix. The offset is mostly within $5 \%$ (except for the outer redshift bin) and approaches unity for $\ell > 10$, which is the smallest multipole that we shall use in our analysis.
This systematic difference is significantly smaller than the Gaussian random error (see Eq.~\ref{sigma}) that we adopt in our study (see Sect.~\ref{Sec:errors}).

Redshift space distortions modify the APS on the same scales as affected by the Limber approximation. To compare the respective amplitude of the two effects we 
show, in the bottom panels of Fig.~\ref{class}, the amplitude of the RSD signal, computed as the ratio between the 2MPZ angular spectra in real and redshift space, as obtained from \texttt{CLASSgal}. The amplitude of the RSD effect is comparable to the systematic error introduced when the Limber approximation is adopted.
From this comparison we conclude that \textit{i}) the Limber approximation in Eq.~(\ref{limber}) provides fair estimates of the real space APS for $\ell\geq 10$, and \textit{ii}) in this $\ell$-range, the APS is not affected by RSD, either in the first and second redshift bins. In the third redshift bin the RSD signal is comparable to the random error, but only below $\ell \sim 10$.  

Following the above results, in order to avoid unnecessary approximations, in our likelihood analysis we shall implement the exact expression for the APS with RSD, despite the computational cost. 

\subsection{The angular power spectrum estimator}\label{sec:est}

In this work we use the estimator of APS introduced by  \citet[][]{1973ApJ...185..413P} (see also \citealt{1973ApJ...185..757H, 1994ApJ...436..443W, PhysRevD.64.083003}), and employed in many analyses, including tomographic ones similar to ours
\citep[e.g.][]{2004MNRAS.351..923B,2007MNRAS.374.1527B,2011MNRAS.412.1669T}.
The estimator implements Eq.~(\ref{almd}) as
\be\label{klt}
\hat{K}^{ij}_{\ell}=\frac{1}{f_{\rm sky}(2\ell+1)}\sum_{m=-\ell}^{m=+\ell} |\hat{a}^{i}_{\ell m}\hat{a}^{* j}_{\ell m}|-\frac{1}{\bar{\sigma}_{i}}\delta^{K}_{ij} \, ,
\ee
where the second term represents the Poisson shot-noise correction. We verified that such a model for the shot-noise is adequate for the 2MPZ catalogue as it matches the angular spectrum
of a random distribution of objects with the same surface density. Comparisons with model predictions use the ensemble average of Eq.~(\ref{klt})
\be \label{kldt}
\langle \hat{K}^{ij}_{\ell} \rangle =\frac{1}{f_{\rm sky}}\sum_{\ell'} R_{\ell \ell'} C^{ij}_{\ell'},
\ee
which includes the mixing matrix $R_{\ell \ell'}$ (Eq.~\ref{mm}). 

The practical implementation of the estimator consists of two steps. First of all we use the \texttt{HealPix} package to estimate the harmonic coefficients of a pixelized galaxy surface density map, 
\be
\hat{a}^{i}_{\ell m}=\Delta \Omega_{p} \sum_{k=1}^{N_{\rm pix}}\lp \frac{\mathcal{N}_{ik}-\bar{\mathcal{N}}_{i}}{\bar{\mathcal{N}}_{i}}\rp  Y^{*}_{\ell m}(\hvr),
\ee
where $\mathcal{N}_{ik}$ is the number of 2MPZ  galaxies in the $k-$th pixel and $\bar{\mathcal{N}}_{i}$ its mean in the $i$-th redshift shell. All the pixels have equal area $\Delta \Omega_{p}$. The resolution matches that of the angular 2MPZ mask and corresponds to $\ell_{\rm max} \simeq 256$. We average the measurements obtained from Eq.~(\ref{klt}) as
\be\label{cdel}
\hat {C}^{ij}_{\Delta \ell}= \frac{\sum_{\ell \in \Delta \ell}(2\ell +1)\hat{K}^{ij}_{\ell}}{\sum_{\ell \in \Delta \ell}(2\ell +1)},
\ee
where we have chosen $\Delta \ell=6$ in order to minimize the number of elements of the covariance matrix, while reducing the effect of the window function by
keeping about $\sim 90\%$ of the original signal in the $\ell$ bin, as discussed in Sect.~\ref{sec:mix}.
The bin-average mixing matrix is computed as
\be\label{dmm}
R_{\Delta \ell, \ell'}=
\frac{(2\ell'+1)}{4\pi}\sum_{\ell''}(2\ell''+1)W_{\ell''}\mathcal{W}_{\Delta \ell,\ell',\ell''},
\ee
where $\mathcal{W}_{\Delta \ell,\ell',\ell''}$ denotes the $3j$-Wigner symbols averaged as in Eq.~(\ref{cdel}).

\begin{figure}
  \vspace{-1cm}
  \includegraphics[width=9cm] {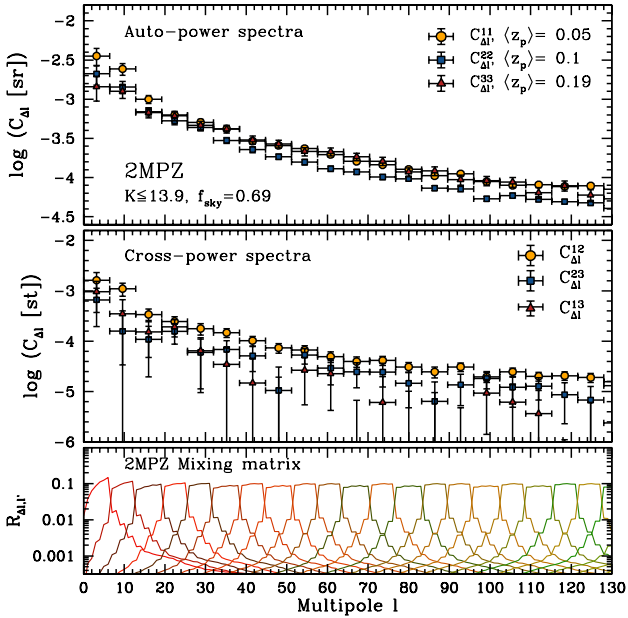}
  \vspace{-2cm}
  \caption{The 2MPZ angular power spectrum in the three \phz\ bins defined in the text. The error bars were derived from the Gaussian approximation, sufficient for our purposes. The upper panel shows the auto-power spectra of the 2MPZ. The middle panel presents the cross-power spectra among the redshift bins. The bottom panel illustrates the elements of the $\ell$-averaged mixing matrix $R_{\Delta \ell \ell'}$ (see Eq.~\ref{dmm}). }\label{cl1}
\end{figure}

Other estimators based on the harmonic decomposition have been used to estimate angular spectra of galaxies \citep[e.g.][]{2004MNRAS.351..923B,2007MNRAS.374.1527B,2011MNRAS.412.1669T}.
We compare one of them with the estimator used here in Appendix \ref{sec:estimator_ap}, observing no significant difference between the two results. There are also alternative approaches to measure the APS from a galaxy sample, such as the maximum likelihood \citep[e.g.][]{2001ApJ...555..547H, 2002ApJ...571..191T, 2004MNRAS.351..923B, 2012ApJ...761...13S,2013MNRAS.428.3487H}. In particular, \citet[][]{2004MNRAS.351..923B} showed that the harmonic analysis (as the one we adopted here) and the maximum likelihood estimator yield estimates of APS that are in good agreement, when applied on samples with large sky coverage, as is the case of 2MPZ. Also, publicly available codes such as \texttt{PolSpice} \citep{2004MNRAS.350..914C} have been implemented to obtain APS in order to perform homogeneity tests in the 2MPZ sample \citep{alonso}. We have developed our own APS code, \texttt{H-GAPS} (\textit{Healpix-based galaxy angular power spectrum}), which we release together with this paper\footnote{\url{https://abalant.wixsite.com/abalan/to-share-1}}.


\section{Results}\label{results}
In this Section we present the main results of the measurement of 2MPZ APS in the three adopted redshift bins, both for auto- and
 cross-power spectra. We then validate them by computing the errors (covariance matrices) using three different approaches.

\subsection{The measurements of the 2MPZ angular power spectrum}

In  the upper panel of Fig.~\ref{cl1} we show the measurements of the $\ell$-binned, angular auto-power spectra of 2MPZ galaxies in three \phz\ bins, illustrated with three different symbols. In the multipole range shown here the signal dominates over the shot-noise error in the first two redshift bins. In the third $z$-bin the shot-noise becomes larger than the signal for $\ell \geq 70$. The middle panel of Fig.~\ref{cl1} shows the angular cross-spectra between galaxies in different bins. Not surprisingly, the amplitude of the cross-spectrum is significantly smaller than that 
of the auto-spectrum, especially in the case of the first vs. third redshift bin (red triangles).
The error bars show Gaussian errors which, as we will show in Sect.~\ref{sec:errcomp}, provide a good estimate of the uncertainties.
The bottom panel shows the elements of the mixing matrix obtained with Eq.~(\ref{dmm}), showing how the signal from a given $\ell-$bin is spread towards neighbouring bins due to partial sky coverage\footnote{A full-sky coverage would lead to bin-averaged mixing matrix given by rectangular functions.}.

Focusing on the auto-spectra, we see that the spectral amplitude decreases from redshift bin 1 to redshift bin 2, and then increases again in redshift bin 3. 
This apparently anomalous behaviour reflects the interplay between the evolution of galaxy clustering and its luminosity dependence in a dataset such as 2MPZ.
Evolution lowers the amplitude of the clustering signal as a function of redshift, provided that the same population of objects is selected. This is basically the 
case when moving from redshift bin $1$ to bin $2$.
The second effects dominates in the third redshift bin in which, because of the flux-limit, the selected 2MPZ galaxies are intrinsically brighter, more biased and, consequently, more clustered than in the first two redshift bins.

The shape of the angular spectrum is well-approximated (in the range $20 \leq \ell \leq 100$ ) by a power-law $C_{\ell}=A\ell^{-\gamma}$. For the $K\leq 13.9$ limit we obtain $A=(4.6\pm 0.8, \, 6\pm 1, \, 2.5\pm 0.6)\times 10^{-2}$ and $\gamma=1.35\pm 0.04, \, 1.51\pm 0.05, \, 1.18\pm 0.06$ in the first, second, and third redshift bin, respectively. Below $\ell=20$ the signal is modulated by the competing effects of RSD and the geometry mask. In Sect.~\ref{sec:mix} we have seen that the amplitude of these systematic effects is significantly smaller than that of the random errors which, on these scales, are rather large. That said, we find no evidence for an excess power on these scales, apart from a steepening at $\ell<15$ which seems to be more prominent in the first redshift shell. The depth of this bin is comparable to that of the sample analysed by \citet[][]{2005MNRAS.361..701F}, that, however, was brighter than ours ($K\leq 12.5$). These authors also detected excess power, but it was located in the range $\ell=[5,30]$, which only partially overlaps with the multipole interval we consider here.

\begin{figure}
  \vspace{-2cm}
  \includegraphics[width=8.5cm]{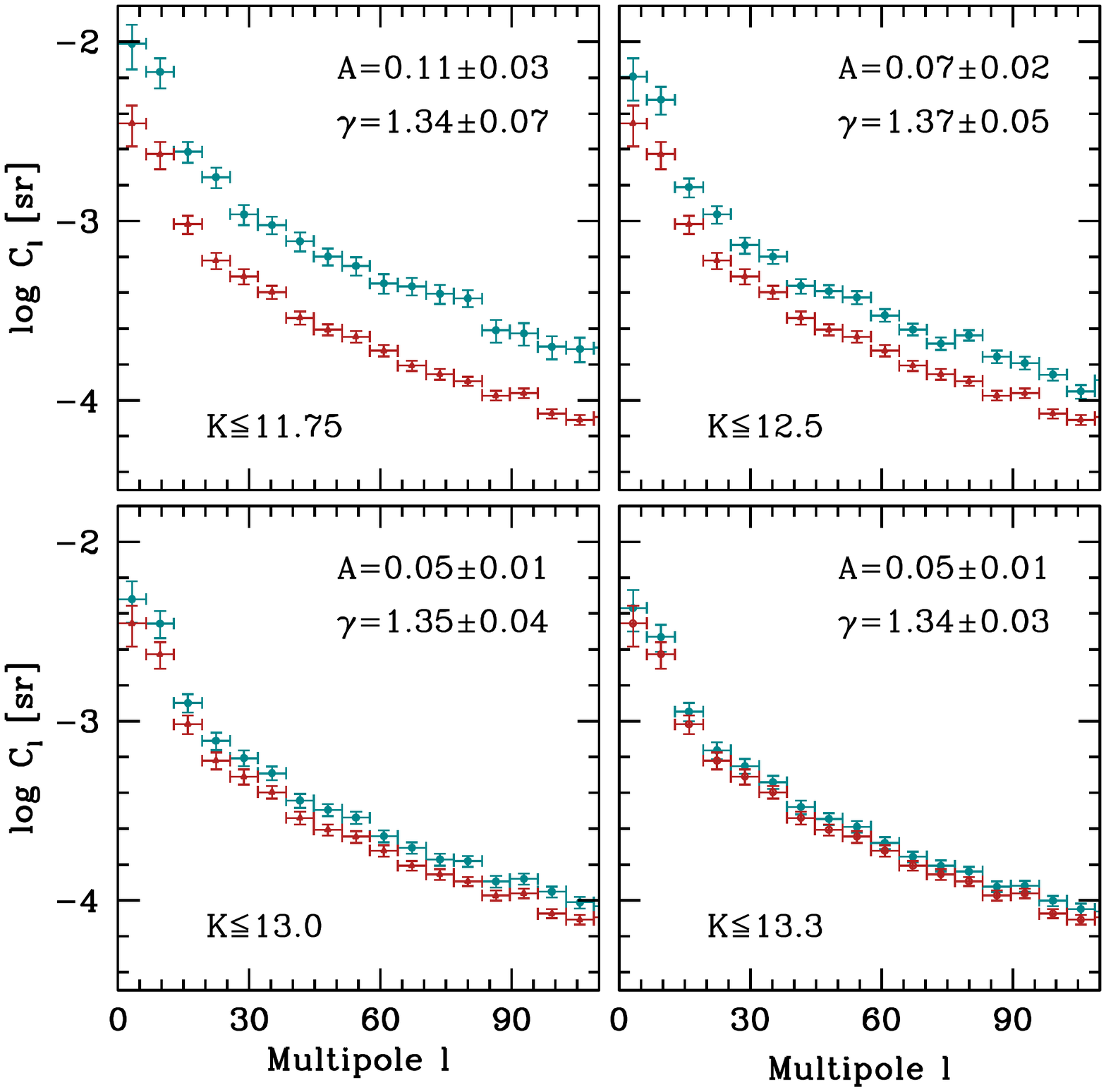}
\vspace{-2cm}
\caption{2MPZ angular power spectrum as a function of the $K$ apparent magnitude cut for galaxies in the first redshift bin, i.e. $z_{\rm p}<0.08$. Red triangles in all the panels show the power spectrum 
computed using all galaxies brighter than the fiducial $K= 13.9$ limit, for comparison. 
The numbers quoted correspond to the parameters of the best fit $C_{\ell}=A \ell^{-\gamma}$, in the range $20\leq \ell \leq 100$.}\label{clmk}
\end{figure}

For a more self-consistent, though still largely qualitative comparison, one should enforce similar flux-cuts to both catalogs. This is the scope of Fig.~\ref{clmk}, in which we show the angular power spectra in the first redshift bin for 2MPZ galaxies selected at different flux cuts, indicated in the panels. The difference in the spectral amplitudes quantifies the effect of the luminosity-dependent bias. The top-right panel compares the angular power spectrum of all 2MPZ galaxies in the first redshift bin (red symbols) with that of galaxies brighter than $K=12.5$, the same cut as in \citet[][]{2005MNRAS.361..701F}. The effect of the cut is to significantly change the amplitude of the spectrum but not the shape. As a result, the excess power is seen at all flux cuts. We conclude that the large power at $\ell<20$ is a robust feature of the 2MPZ spectrum that partially overlaps with the excess power detected by \citet[][]{2005MNRAS.361..701F}. Whether or not this represents an anomaly with respect to the model predictions will be discussed in Sect.~\ref{sec:ind_red_bin}.


\subsection{Error analysis}
\label{Sec:errors}
Most of the previous APS analyses of \phz\ samples \citep[e.g.][]{2004MNRAS.351..923B,2011MNRAS.412.1669T,alonso} have assumed Gaussian errors, showing that they were adequate for the level of accuracy required in those studies. Similarly, we now assess the goodness of the Gaussian hypothesis for a sample like 2MPZ and compare
it with two alternative, and arguably more reliable, error estimates: those obtained from the 2MPZ mock catalogues described in Sect.~\ref{smocks}, and those derived from the so-called jackknife technique.

\subsubsection{Gaussian Errors}\label{sec:cova}
Under the assumption that, in the $i-$th redshift bin, the spherical harmonic coefficients $a^{i}_{\ell m}$ are Gaussian random distributed variables, the covariance matrix of the angular cross-power spectrum is diagonal, with a variance given by \citep[e.g.][]{1997PhRvD..55.7368K}:
\be\label{sigma}
\sigma^{(ij)}_{\ell}=\sqrt{\frac{2}{(2\ell+1)f_{\rm sky}}}\left[(C^{ij}_{\ell})^{2}+\lp C^{(i)}_{\ell}+S_{i} \rp\lp  C^{(j)}_{\ell}+S_{j} \rp \right]^{1/2},
\ee
for $i\neq j$, where $S_{i}$ is the shot-noise of the APS measured in the $i$-th redshift bin. The variance for the auto-power spectrum is given by \citep[e.g.][]{2003moco.book.....D}
\be\label{sigma}
\sigma^{i}_{\ell}=\sqrt{\frac{2}{(2\ell+1)f_{\rm sky}}}\lp C^{i}_{\ell}+S_{i} \rp.
\ee

\begin{figure*}
  \vspace{-2cm}
  \includegraphics[width=16cm]{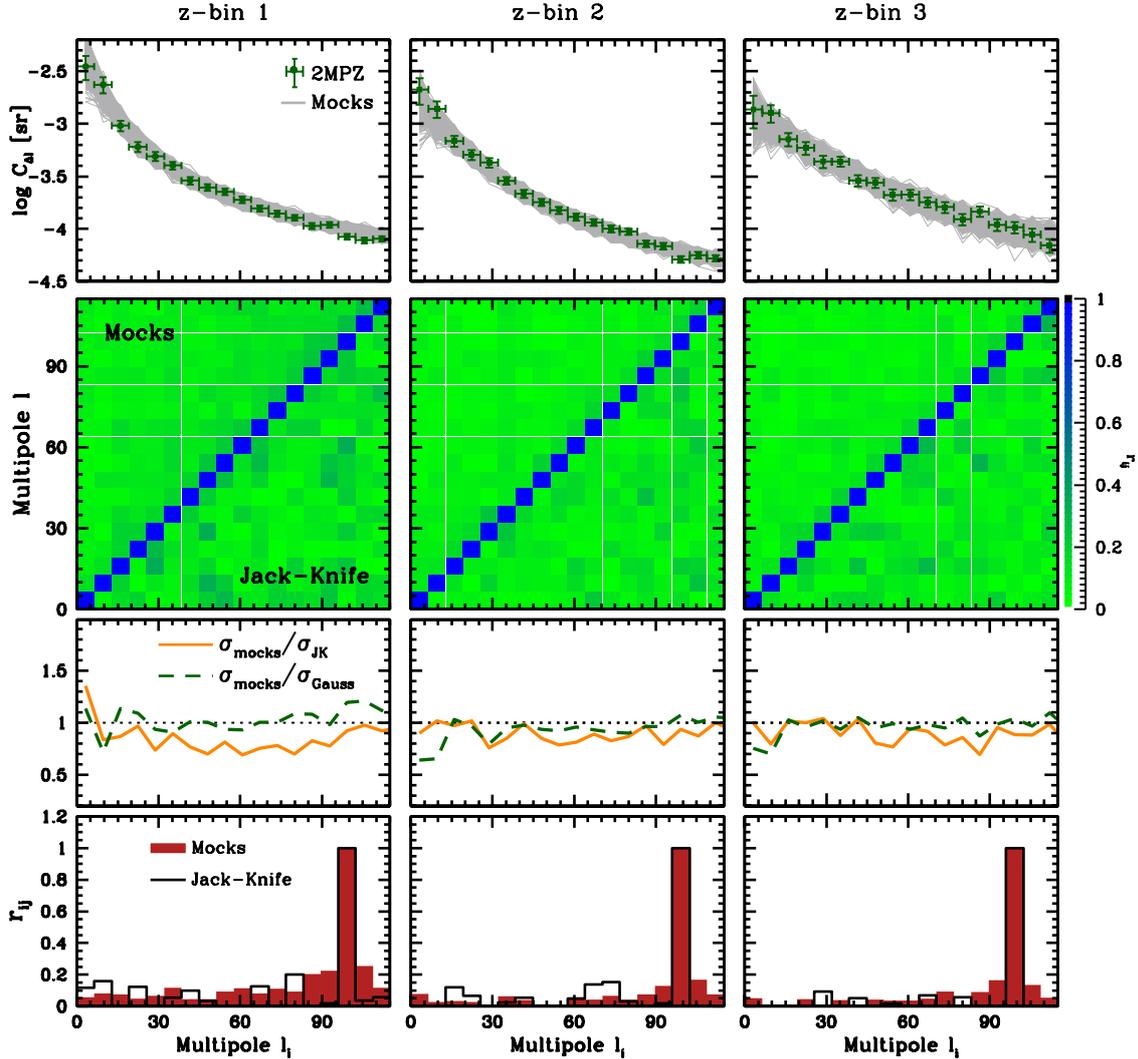}
\vspace{-4cm}
  \caption{2MPZ angular power spectrum error comparison. Top panels: 2MPZ angular spectra (green dots) vs. individual mock spectra (grey curves). Vertical bars represent Gaussian errors.
  Second row: Covariance matrix elements estimated from the mocks (upper half) and from jackknife (lower half), both normalized to their diagonal elements. The colour code 
  represents the amplitude. Third row: comparison between diagonal elements: mocks vs. JK (orange solid) and mocks vs. Gaussian (green dashed).
  Bottom panels: histograms representing the amplitude of the correlation matrix elements centred at $\ell =100$. JK (empty histograms) vs. mocks (filled histograms).
  Results in the three columns refer to the three 2MPZ redshift bins indicated in the labels.}\label{cl_cor}
\end{figure*}


\subsubsection{Covariant errors from the 2MPZ mock catalogues}\label{sec:errormocks}

A better estimate of the errors which also accounts for their covariance can be obtained by exploiting the mock 2MPZ catalogues described in Sect.~\ref{smocks}. In this case the accuracy of the error estimate depends on the number of available mocks and their similarity to the real sample.

The relation between the accuracy and the number of mocks $N_{M}$ is not trivial and depends on the number of free parameters in the analysis, $N_{P}$, and the number of bins in which the clustering measurement is performed, $N_{K}$. If $\sigma^{2}_{0}$ are the ideal values of the diagonal element of a covariance matrix obtained from an arbitrary large number of mock catalogues, then the additional variance $\sigma^{2}_{\rm add}$ induced by using a limited number $N_{M}$ of mocks to estimate the covariance matrix is $\sigma^{2}_{\rm add}/\sigma^{2}_{0}\approx (N_{K}-N_{P})/(N_{M}-N_{K})$ \cite[e.g.][]{2013PhRvD..88f3537D}. In our case we use $N_{K}\sim 10$ $\ell$-bins to constrain $N_{P}=4$ cosmological parameters. Therefore we need $\gtrsim 700$ mocks in order to guarantee that the additional variance is below $\sim 1 \%$.

The similarity between mock and real samples has been discussed in Sect.~\ref{smocks}. Here we stress the fact that that in the mocks the APS multipoles are all independent, despite the fact that a lognormal PDF is assumed. To estimate  covariant errors we compute the binned angular spectra in the three redshift bins of each mock and compute the covariance matrix as:
\be
\mathcal{C}_{\ell \ell'}=\frac{1}{N_{M}-1}\sum_{j=1}^{N_{K}}\lp \tilde{C}^{(j)}_{\ell}-\bar{\tilde{C}}_{\ell}\rp \lp  \tilde{C}^{(j)}_{\ell'}-\bar{\tilde{C}}_{\ell'}\rp, 
\ee
where $N_{M}=1000$. $\bar{\tilde{C}}_{\ell}$ denotes the sample mean.


\subsubsection{Jackknife errors}
The jackknife (JK) resampling \citep{1958} techniques allows one to estimate random errors from the dataset itself, with no need to use mock catalogues. This approach has been extensively applied to multiple galaxy clustering analyses \cite[see e.g.][]{2007MNRAS.381.1347C, 2009MNRAS.396...19N, 2011MNRAS.418.2435N,2016arXiv160600233E}.
Its implementation for a 2D sample consists of dividing the observed sky into non-overlapping, equal-area regions and computing the relevant quantity (APS for the present work) after removing one of such regions at a time. The various regions are represented by a set of low resolution $\tilde{N}_{\rm side}$ \texttt{Healpix} pixels (patches hereafter). Because of the 2MPZ geometry mask, the number of unmasked small pixels (used for the clustering analysis) varies from patch to patch. Therefore, in order to have a minimal number of JK patches $N_{JK}$, we have only considered those in which the scatter in the number of unmasked pixels deviates by less than $20\%$ from the mean. After measuring the APS in each of these $N_{s}={{N_{JK}}\choose{d}}$ JK replicates, where $d$ is the number of masked-out sky patches, 
we compute the error covariance matrix as
\be\label{jkcov}
\mathcal{C}_{\ell \ell'}=\frac{N_{JK}}{N_{s}d}\sum_{j=1}^{N_{s}}\lp C^{(j)}_{\ell}-\bar{C}_{\ell}\rp \lp  C^{(j)}_{\ell'}-\bar{C}_{\ell'}\rp.
\ee
where $\bar{C}_{\ell}$ is the mean among the $N_{s}$ replicates. In general, the results depend on the patch size, set by the resolution $\tilde{N}_{\rm side}$, and the number of masked-out regions $d$. We have explored 
different combinations of $\tilde{N}_{\rm side}$ and $d$ and found that the mean of the $N_{s}$ JK replicates $\bar{C}_{\ell}$, and the diagonal elements of the associated covariance matrix (Eq.~\ref{jkcov}) obtained from the configuration ($\tilde{N}_{\rm side}=4$, $d=1$)
 agree, within $\sim 1\%$ and $\sim 10\%$ respectively, with the same quantities obtained from the ensemble of mocks. With these parameters we obtain a set of $N_{s}=N_{\rm JK}=119$ JK replicates.


\subsection{Error comparison}\label{sec:errcomp}

Figure \ref{cl_cor} summarizes and compares the results of the various error estimates. We focus here on the angular auto-spectra.
The three columns show the results obtained in the three redshift  bins.
The top panels compare the measured APS of 2MPZ galaxies (green dots) with those obtained from the 1000 2MPZ mock catalogues (overlapping grey
curves). The angular spectra of the mocks are in good agreement with those of the real 2MPZ catalogue, demonstrating that the procedure described in Sect.~\ref{smocks}, based on a log-normal probability distribution, generates realistic mocks. The scatter among the mocks also matches the Gaussian error bars.

 The plots in the second row of Fig.~\ref{cl_cor} compare the off-diagonal elements of the covariance matrices computed using the mock catalogues (the upper half of each panel) and the jackknife method (lower half). Each bin represents one element of the matrix, colour-coded according to its amplitude, normalized to the diagonal elements.
In both cases the amplitude of the off-diagonal elements is less than 20$\%$ of the diagonal elements.
Off-diagonal terms arise from the mode-coupling induced by the geometry mask and by the nonlinear evolution. The latter is ignored in the mock catalogues.
This partly explains why these terms are larger in the JK matrices than in the mock matrices. Another source of mismatch comes from the fact that JK error estimate is less accurate than that obtained from the 1000 mocks \citep[e.g.][]{2009MNRAS.396...19N}.

The third row of Fig.~\ref{cl_cor} compares the amplitude of the diagonal errors computed using the three methods. 
The amplitude of the Gaussian errors is very similar to that of the diagonal errors obtained from the mocks, except at very small $\ell$ values (green dashed curves).
This result is consistent with the small amplitude of the off-diagonal elements which, in turns, is a manifestation of the large sky coverage of the 2MPZ catalogue.
The orange solid curve shows that, instead, JK errors are systematically larger than the ones obtained from the mocks. The effect is stronger in the 
first redshift bin, where the amplitude of the mismatch can be as large as $30\%$, reducing to $10-15\%$ at higher redshift.
This redshift dependence is not surprising and mainly reflects the impact of nonlinear effects which, at small redshifts, can propagate to large angular scales.

It is worth noticing that the larger amplitude of the JK error is contributed by objects in a limited number of sky patches in which the clustering amplitude is significantly
larger than the mean signal. We plan to investigate deeper the significance of these effects and the properties of 2MPZ galaxies residing in these areas in a follow-up paper \citep[see e.g.][for a related approach]{2016MNRAS.460..256A}.

In the bottom panels of Fig.~\ref{cl_cor} we compare the elements of the correlation matrices for the bin centred at $\ell =100$ for the JK (solid line histograms) and the 
2MPZ mock errors (filled, red histograms). The amplitude of the terms which are far from the diagonal is larger in the JK case, whereas terms close to the diagonal
are larger in the mock case. 

These results show that differences in the random errors computed using different methods are smaller than the 
error amplitudes, and that off-diagonal elements are small.
Therefore, in the likelihood analysis, we assume random Gaussian errors with no covariance.
We demonstrate in Appendix \ref{ap:err} that this choice does not have an impact on the results of the likelihood analysis.


\section{Likelihood analysis}\label{sec:lik}

In this Section we compare the measured 2MPZ angular auto- and cross-spectra with the theoretical predictions of the $\Lambda$CDM model to 
estimate a set of cosmological parameters $\theta$. To do this, we sample the posterior conditional probability of $\theta$ given the measured angular spectrum $\hat{C}^{ij}_{\Delta \ell}$, 
$\mathcal{P}(\theta|\hat{C}^{ij}_{\Delta \ell})$, using a  MonteCarlo Markov-Chain approach.
The Bayes theorem guarantees that $\mathcal{P}(\theta|\hat{C}^{ij}_{\ell})\propto  \mathcal{P}(\theta)\mathcal{L}(\hat{C}^{ij}_{\Delta \ell}|\theta)$. 
For a flat prior  $\mathcal{P}(\theta)$ we sample the likelihood which is assumed to be Gaussian
$\mathcal{L}(\hat{C}^{ij}_{\Delta \ell}|\theta) \propto {\rm e}^{-\chi_{ij}^{2}/2}$, with
\be
\chi_{ij}^{2}=\lp C^{ij}_{\Delta \ell}(\theta)-\hat{C}^{ij}_{\Delta \ell} \rp \mathcal{\textbf{C}}^{-1}\lp C^{ij}_{\Delta \ell'}(\theta)-\hat{C}^{ij}_{\Delta \ell'}\rp,
\ee
where $C^{ij}_{\Delta \ell}(\theta)$ is the model power spectrum of Sect.~\ref{sec:power1}, which includes the effect of the mixing matrix, and $\mathcal{\textbf{C}}^{-1}$ is the inverse of the covariance matrix of Sect.~\ref{sec:cova}. 
Following the conclusions of that Section, we ignore off-diagonal terms.

To sample the posterior probability we use the publicly available code \texttt{MontePython} \citep[][]{Audren:2012wb}. 
To combine measurements from different bins we simply multiply the respective posteriors, i.e. we assume no correlation among the redshift bins.
Finally, to obtain the 2D and 1D confidence intervals we marginalize the posterior over all the other parameters.

\begin{figure}
  \vspace{-2.5cm}
  \hspace{-0.5cm}
  \includegraphics[width=11.5cm]{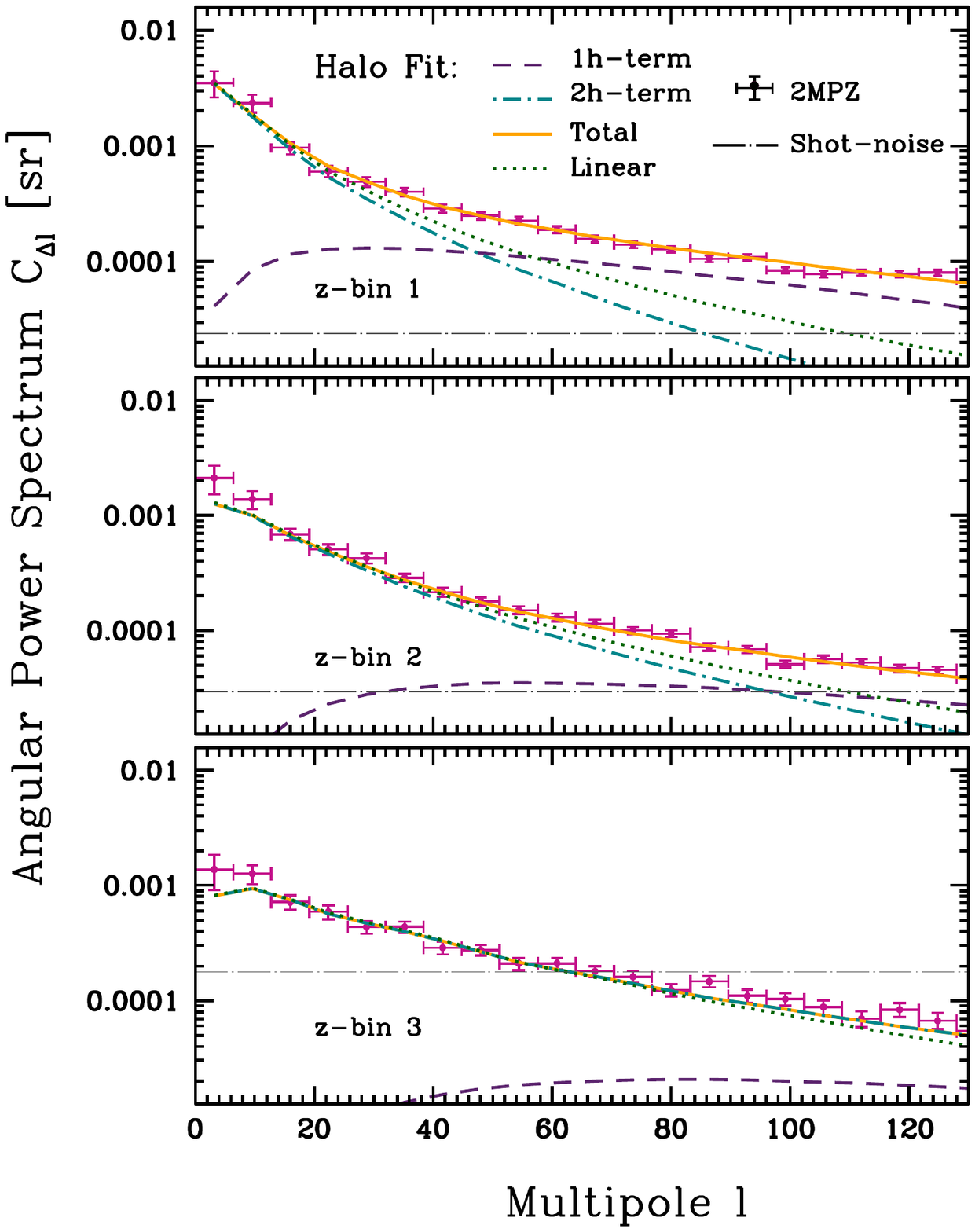}
  \vspace{-3.5cm}
  \caption{The 2MPZ binned angular auto-power spectrum (red dots with Gaussian error bars) in three bins of increasing redshift (from top to bottom).
  The orange continuous curve is the  \texttt{Halo-fit} model spectrum and its 1-halo and 2-halo contributions (dashed and dot-dashed curves). This model assumes the fiducial cosmology. The linear model (dotted curve) is also shown for reference.
  Model spectra have been boosted up by linear bias factors, as discussed in the text. The horizontal 
  long-dashed-dotted curve indicates the shot-noise level in each redshift bin.}\label{cl_mod}
\end{figure}

We focus on the same cosmological parameters as determined in previous tomographic analyses, namely, the mass density parameter of the dark matter component $\Omega_{\rm cdm} \in  [0,0.7]$, the baryon energy density parameter $\Omega_{\rm b}\in [0,0.09]$, the amplitude of the primordial power spectrum (at a pivot scale of $0.05$ $h$ Mpc$^{-1}$), $10^{9}A_{S}\in[0.1,10]$ and the linear galaxy bias in each redshift bin $b_{i}\in[0.1,10]$. 
The values in the parentheses are ranges of the (flat) priors. We map this parameter space into the set $\{f_{\rm b},\Omega_{\rm mat},\sigma_{8}, b\}$ where $\Omega_{\rm mat}=\Omega_{\rm cdm}+\Omega_{\rm b}$ is the total matter energy density parameter, $f_{b}=\Omega_{\rm b}/\Omega_{\rm mat}$ is the baryon fraction, and $\sigma_{8}$ is the \textit{rms} of the matter distribution on spheres of radius $8$ Mpc $h^{-1}$ (at $z=0$), which is related to $A_{S}$ and normalizes the linear power spectrum \citep[see e.g.][]{2009ApJS..180..330K}. Except for the galaxy bias, all parameters are specified at $z=0$. 

To compare model and data we need to indicate the multipole range considered in the analysis.
We set the minimum value at $\ell=10$ to minimize the impact of the systematic errors induced by the geometry mask, which we discuss
in details in Appendix \ref{sec:mask}.
For the maximum $\ell$ we choose a conservative value that accounts for the impact of both the map resolution (i.e. the pixel size) and
that of shot-noise. The effect of pixel size is redshift-independent and, as shown in Appendix~\ref{sec:mask}, becomes important  for $\ell \sim100$.
The impact of shot-noise depends on the redshift due to the flux-limited nature of the sample and can be appreciated in Fig.~\ref{cl_mod} by comparing the shot-noise level (horizontal long-dashed lines) with the measured 2MPZ APS (points with Gaussian error bars).

We point out that in the $\ell$-ranges considered here, departures from the linear model are significant in the first two redshift bins. This can be approximately justified by Fig.~\ref{cl_mod}, where the orange solid curves in each panel show the model of the APS for the fiducial cosmological setup, for the three redshift bins. This model has been obtained using  \texttt{CLASSgal} and includes \texttt{Halo-Fit} \citep[][with the 1-halo and 2-halo terms represented by the dashed and the dot-dashed curves, respectively]{2003MNRAS.341.1311S, 2012ApJ...761..152T}
to account for non-linear evolution of the underlying dark matter. The linear APS (computed with the same set of fiducial parameters) is also plotted for reference (dotted curve). Model spectra have been boosted up to match the amplitude of the measured ones at $\ell \sim 20$. 

We want to highlight the fact that at the small angular scales we are able to probe before shot-noise domination (i.e. $\ell\sim 100$) and the redshift range covered by our analysis, even if we account for the non-linear clustering of the dark matter, a constant galaxy bias is an inaccurate approach to model galaxy clustering \cite[e.g.][]{2007PhRvD..75f3512S}. In other words, pushing the analysis until $\ell=100$ would demand increasing the number of parameters to account for galaxy bias. We therefore decided to set a more conservative value of $\ell_{\rm MAX}=70$ for the cosmological analysis. This angular scale represents a minimal physical separation of $\sim 15,\,\, 25$ and $40$ Mpc $h^{-1}$ for the first, second and third redshift bins, respectively.

Note that by using \texttt{Halo-Fit} to model the underlying matter power spectrum, we can attempt to generate individual estimates on the parameters $\sigma_{8}$ and $b$, which are degenerated in the linear regime. Finally, as commented in Sect.~\ref{rsddd}, and in order to be as general as possible, our APS model includes the effects of RSD.

Finally, the plots show that the model provides a good fit also below $\ell=15$, i.e, on the scales where the 2MPZ APS steepens, as discussed in Sect.~\ref{results}. The good match between the model and data indicates that the steepening of the APS at large angular scales is not anomalous. Instead, it is in good agreement with $\Lambda$CDM predictions. We conclude that we find no support to the claim of excess power on large scales by e.g. \cite{2005MNRAS.361..701F}.


\begin{figure}
  \vspace{-0.5cm}
  \includegraphics[width=9.5cm]{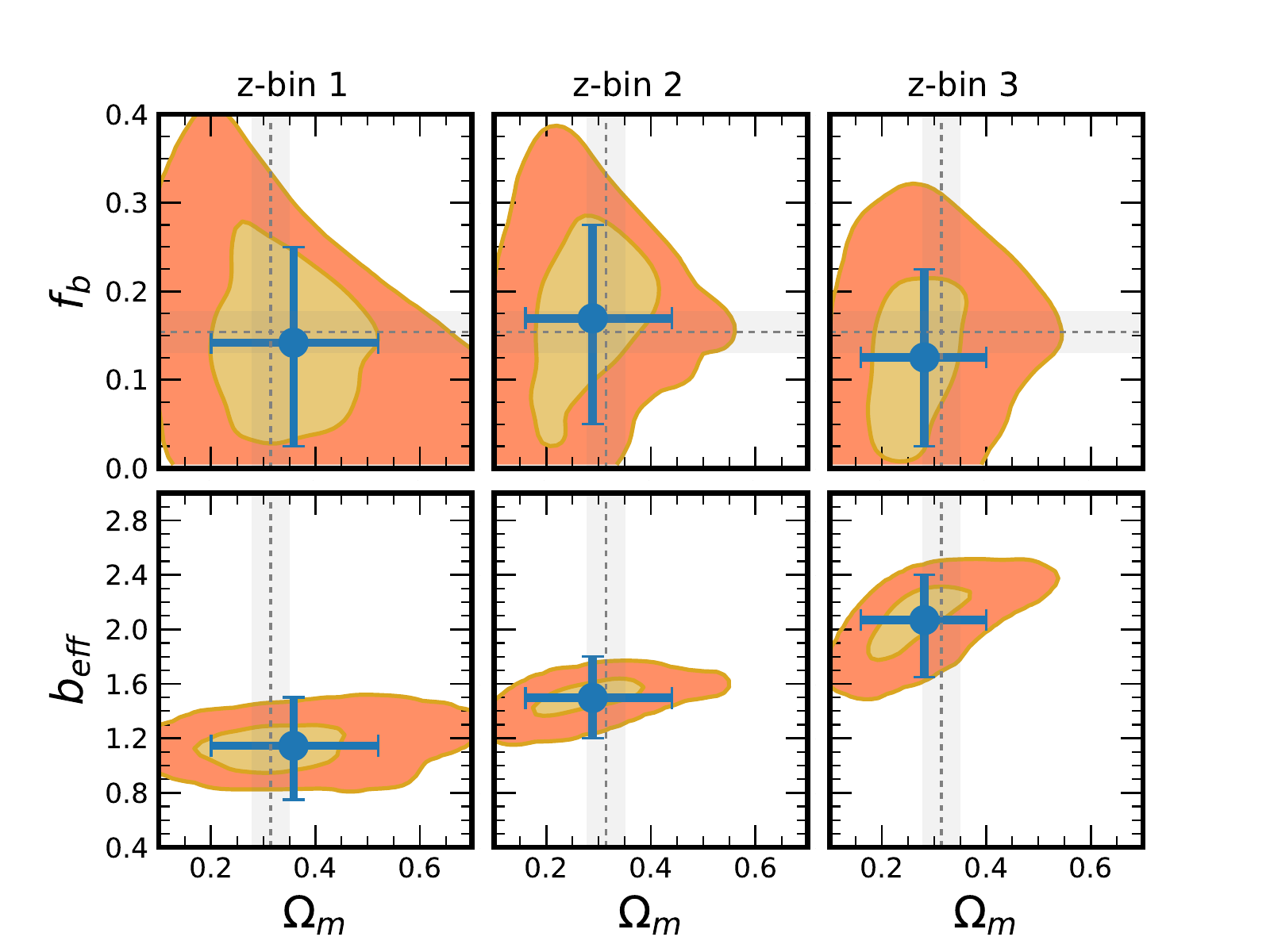}
  \caption{
  $68\%$ and $99\%$ confidence contours for the parameters $f_{b}$ and $\Omega_{\rm m}$, derived for $z=0$ from 2MPZ auto-power spectra in the three
  redshift bins (top panels),  obtained after marginalizing over $b(z)$ and $\sigma_{8}$. 
  Bottom panels show the same confidence contours for $b(z)\sigma_{8}$ and $\Omega_{\rm m}$. Blue dots and error bars indicate the best fit values and their $68\%$ confidence intervals in each parameter, obtained after marginalizing over the rest of varied parameters. Dashed lines and grey bands: measurements and 1-$\sigma$ errors from the analysis of the Planck mission.}\label{cl_cosmo_2mpz}
\end{figure}

\subsection{Individual redshift bins}
\label{sec:ind_red_bin}
In this Section we estimate the cosmological parameters $f_{\rm b}$ and $\Omega_{\rm m}$ that determine the shape of the angular spectra, and the combination
$\sigma_{8}b_i (z=z_i)$ that represents the linear {\it rms} galaxy density fluctuation in the $i-$th redshift bin and sets its amplitude. All the other cosmological parameters are fixed at their fiducial values.
The upper panels of Fig.\ref{cl_cosmo_2mpz} show the $68\%$ and $99\%$ confidence regions in the $\{f_{\rm b},  \Omega_{\rm m} \}$ plane obtained after
marginalizing over $\sigma_{8} b_i$.
The blue dot represents the best fit values and the error bars show the $68\%$ confidence interval on each parameter after marginalizing over the other.
These values are listed in the first two columns of Table \ref{cvalues}. Dashed lines with grey bands illustrate the fiducial parameter values with their $1\sigma$ errors.

Our results agree with those obtained by \citet{2007MNRAS.374.1527B} and \citet{2011MNRAS.412.1669T} who performed a similar, tomographic analysis at larger redshift using SDSS-based MegaZ-DR4 and MegaZ-DR7 catalogues of LRGs, respectively.
Our errors are, however, about twice as large as theirs. This difference, which quantifies the difficulty in carrying out a tomographic analysis in the local Universe, has several causes.
First, 2MPZ is wider than SDSS but the galaxy surface density of the former ($\sim 24$ galaxies per deg$^{2}$) is approximately 3 times smaller than in the LRG sample. As a consequence, shot-noise affects larger angular scales, 
especially in the outer redshift bin of the survey where the galaxy number density drops quickly. Second, non-linear effects in both the underlying dynamics and galaxy 
evolution processes also affect larger scales in the local Universe. Finally, 2MPZ galaxies are significantly less biased, and therefore less clustered, than LRGs. 
The net result is a significant reduction both in the $\ell$-range useful for the likelihood analysis and in the clustering amplitude with respect to the analogous studies based on 
SDSS material. The corresponding errors on the measured cosmological parameters are, therefore, significantly larger.

Nevertheless, the fact that the measured parameters are in the right ballpark is encouraging. This is clear from the comparison with the Planck results \citep[e.g.][]{2014A&A...571A..16P}, also shown in Fig.~\ref{cl_cosmo_2mpz} (dashed lines with error bands).

\begin{figure}
\includegraphics[width=8cm]{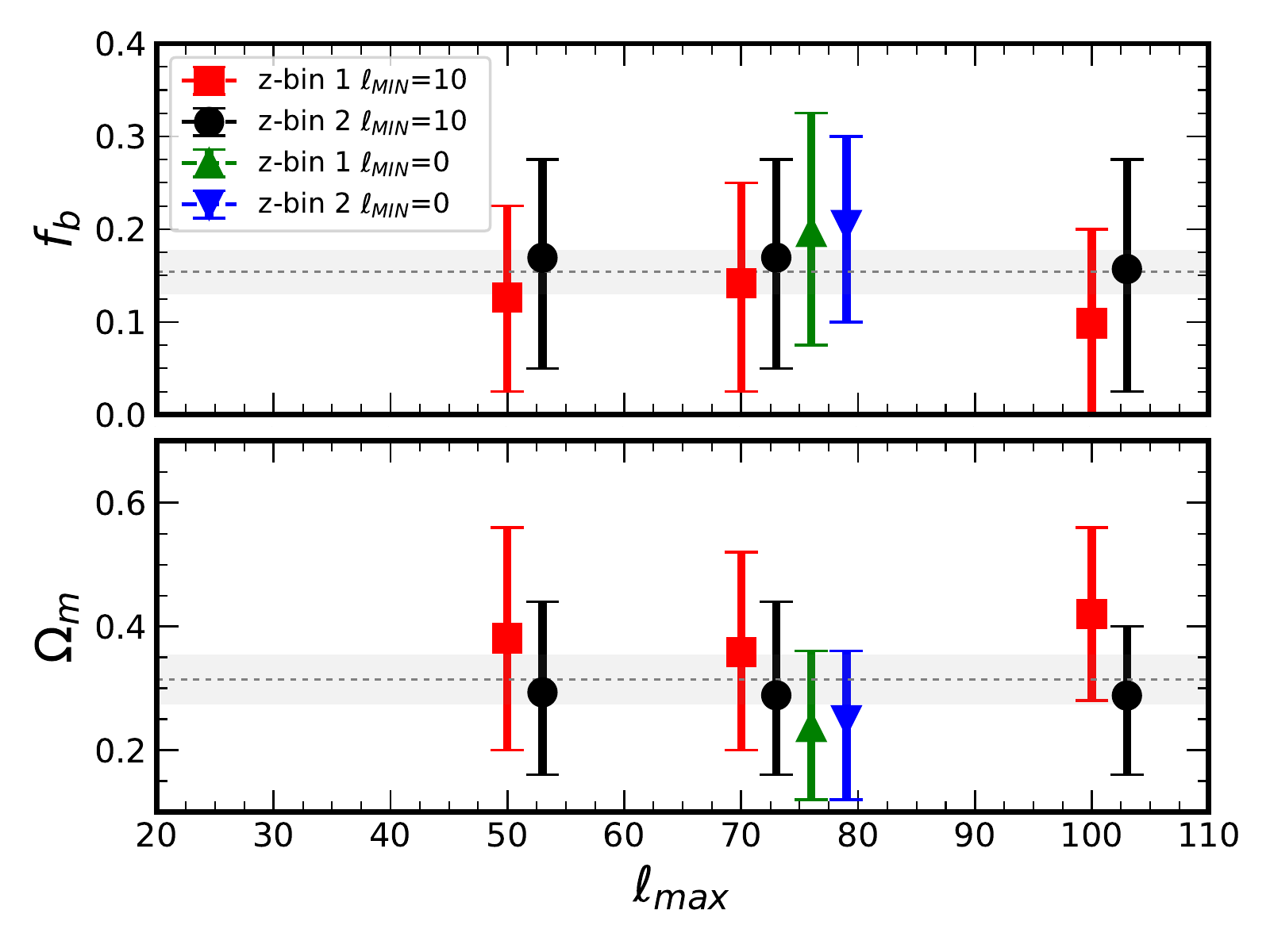}
\caption{Mean values of the parameters $\Omega_{\rm m}$ and $f_{\rm b}$ with their $68\%$ confidence intervals obtained from auto-power spectra in the first (red squares) and second (black circles) redshift bins, as a function of the maximum scale $\ell_{\rm MAX}$ used in the likelihood analysis.  The values from the second redshift bin have been placed at $\ell_{\rm MAX}+3$. Triangles show the results obtained by extending the analysis to the range $[0,70]$ for the first (green triangles placed at $\ell_{\rm MAX}+6$) and the second redshift bin (blue triangles placed at $\ell_{\rm MAX}+9$). The horizontal line and the shaded area represents respectively the Planck values and their $1\sigma$ error bars. }
\label{cl_lmax}
\vspace{-0.5cm}
\end{figure}

The sharp, upper diagonal cutoff in the $99\%$ confidence contour in the $\Omega_{\rm m}$-$f_{\rm b}$ plane of Fig.~\ref{cl_cosmo_2mpz}
  is an artifact that reflects the upper limit that we set on the prior $\Omega_{\rm b} = 0.09$. 
This very generous upper limit, considering the errors in the current measurements of the baryon density, is driven by the consideration that \texttt{CLASSgal}-generated APS models are less accurate for larger $\Omega_{\rm b}$ values. We tested the impact of relaxing this constraint and found that allowing for a larger $\Omega_{\rm b}$ 
broadens the contour and a secondary likelihood peak appears at $\Omega_{\rm b}\sim 0.2$
and $f_{\rm b} > 0.3$.  We regard this second solution as unphysical and decided to stick to our choice of a maximum
 $\Omega_{\rm b}=0.09$.

The three bottom panels of Fig.~\ref{cl_cosmo_2mpz} show the 2D confidence ($68\%$ and $99\%$) regions for the set of parameters $\{\Omega_{\rm m},\,  b_{\rm eff} \}$ (marginalized over $f_b$, for all three redshift bins) where 
\be
b_{\rm eff,i} \equiv \frac{b_i\sigma_{8}}{\sigma^{\rm CMB}_{8}},
\ee
with $\sigma^{\rm CMB}_{8}$ is the {\it rms} mass density parameter obtained by \citet[][]{2014A&A...571A..16P}. The parameter $b_{\rm eff}$ represents the effective linear bias of 2MPZ galaxies brighter than the survey flux limit. In this definition we ignore the weak evolution of $\sigma_{8}$ in the redshift range explored.
The effective bias increases significantly with the redshift, whereas the mass density parameter is in agreement with the Planck value (vertical strip). 

The behaviour of these contours as a function of the redshift bin is as expected and reflects the different bias factors of 
2MPZ  galaxies in the three redshift shells, as discussed in Sect.~\ref{results}. The best fit values for the effective linear bias parameters $b_{\rm eff}$ are listed in Table \ref{cvalues} together with their $68\%$ confidence interval.
The relative errors are in the range $20-30\%$, to be compared with typical $10\%$ errors in the estimate of the LRG galaxies 
obtained by \citet{2011MNRAS.412.1669T}. 
Our results are also in good agreement with the 2MPZ galaxy linear bias parameters obtained by cross-correlating galaxy catalogues with CMB Planck maps 
to search for the integrated Sachs Wolfe effect \citep[][]{2017arXiv171003238S}.


\subsection{Robustness to the choice of $\ell$-range}\label{sec:lrange}

We have tested the robustness of our result to the choice of the $\ell$-range considered in the APS analysis. We performed two different sets of tests.
First, we fixed $\ell_{\rm MIN}$ to its fiducial value ($\ell_{\rm MIN}=10$) and changed $\ell_{\rm MAX}$. The goal was to assess the impact of nonlinear and shot-noise effects by pushing the analysis to smaller angular scales.
Figure \ref{cl_lmax} shows the estimated value of $f_{\rm b}$ (top) and  $\Omega_{\rm m}$ (bottom) as a function of $\ell_{\rm MAX}$.
The results do not change significantly (i.e. within the $1-\sigma$ error bars) with respect to the fiducial case $\ell_{\rm MAX}=70$. In particular, results in the second redshift bin (black dots) are remarkably robust to $\ell_{\rm MAX}$. In the first bin (red squares), pushing the analysis to $\ell_{\rm MAX}=100$ reduces the size of random errors by $\sim 20\%$ but modifies the best fit values of both parameters.
We interpret this result as an indication that, in this case, nonlinear effects do play a role and bias our results.
For this reason we chose to set $\ell_{\rm MAX}=70$ in the analysis. As for the third bin, we did not explore the case $\ell_{\rm MAX}=100$ since
that regime is shot-noise dominated and found that setting $\ell_{\rm MAX}=50$ has the only effect to increase random errors.

In the second test we set $\ell_{\rm MAX}=70$ and extend the analysis down to the first $\ell$-bin (containing modes in the range $\ell\in (0,6)$). The results are shown in the same plot for both the first and the second \phz\ bins (green and blue triangles). Although we notice that including large scale modes induces a shift in the mean of the posterior distributions towards lower values of $\Omega_{\rm m}$ (high values of $f_{\rm b}$), the constrained values are consistent within $1-\sigma$ with the fiducial value $\ell_{\rm MIN}=10$.

\begin{table*}
\center
\begin{tabular}{|c|c|c|c|c|c|c|c|c|c|}
\hline \hline  \\
&\multicolumn{3}{|c|}{\textbf{Auto-power spectra}} & \multicolumn{4}{|c|}{\textbf{Combined auto-power spectra}} &  \multicolumn{2}{|c|}{\textbf{Adding cross-power spectra}} \\ \cline{2-4} \cline{5-8} \cline{9-10}
                &                            &                        &                        &                        &                       &                        &                       &                       &  \\

\phz\ bin & {\blue{$1$}}  &  {\blue{$2$}} & {\blue{$3$}}  &  {\blue{$1\, \&\,2$}} &  {\blue{$2\,\&\,3$}}  &   {\blue{$1\,\&\,3$}} &  {\blue{$1\,\&2\,\&\,3$}} &  {\blue{$ (1\times 2)\, \&$}}  &  {\blue{$ (1\times 2)\, \&\, (2\times 3)\,\& $}}  \\
combination(s) & $\langle z_{\rm p} \rangle = 0.05$           &      $\langle z_{\rm p} \rangle = 0.1$        &       $\langle z_{\rm p} \rangle =0.19$      &                       &                       &                       &                           &  {\blue{$1\,\&2\,\&3$}}  &  {\blue{$1\,\&2\,\&3$}}             \\ 
\hline

                &                            &                        &                        &                        &                       &                        &                       &                       &  \\
{\blue{$f_{b}$}}          &     $0.14^{+0.10}_{-0.11}$ &  $0.17^{+0.10}_{-0.12}$ & $0.12^{+0.10}_{-0.10}$   & $0.18^{+0.10}_{-0.10}$  & $0.14^{+0.08}_{-0.10}$ & $0.14^{+0.08}_{-0.10}$  &   $0.14^{+0.07}_{-0.08}$  &   $0.14^{+0.09}_{-0.08}$ &  $0.15^{+0.09}_{-0.07}$  \\
                &                            &                        &                        &                        &                       &                        &                       &                        &\\
 {\blue{$\Omega_{\rm m}$}}  &   $0.36^{+0.16}_{-0.16}$    &  $0.29^{+0.15}_{-0.13}$ & $0.28^{+0.12}_{-0.12}$  & $0.27^{+0.08}_{-0.08}$ & $0.31^{+0.08}_{-0.07}$ & $0.31^{+0.08}_{-0.08}$  &   $0.29^{+0.06}_{-0.06}$ & $0.30^{+0.07}_{-0.07}$&  $0.30^{+0.06}_{-0.06}$ \\
                &                            &                        &                        &                        &                       &                        &                       &                        &\\
{\blue{$b_{\rm eff, i}$}}     &  $1.14^{+0.35}_{-0.40}$   &  $1.49^{+0.30}_{-0.30}$  & $2.07^{+0.33}_{-0.48}$   &  &  &   &  &    &   \\

                &                            &                        &                        &                        &                       &                        &                       &                        &\\

\hline \hline
\end{tabular}
\caption{Best-fit values of the relevant cosmological parameters and their $68\%$ confidence intervals obtained by: i) performing auto-correlation analyses in each \phz\ bin (first three columns),
  ii) combining the results of different bins (columns 4 to 7) and iii) from the cross-correlation analysis in different bins (last two columns).}
\label{cvalues}
\end{table*}


\begin{figure}
  \includegraphics[width=8cm]{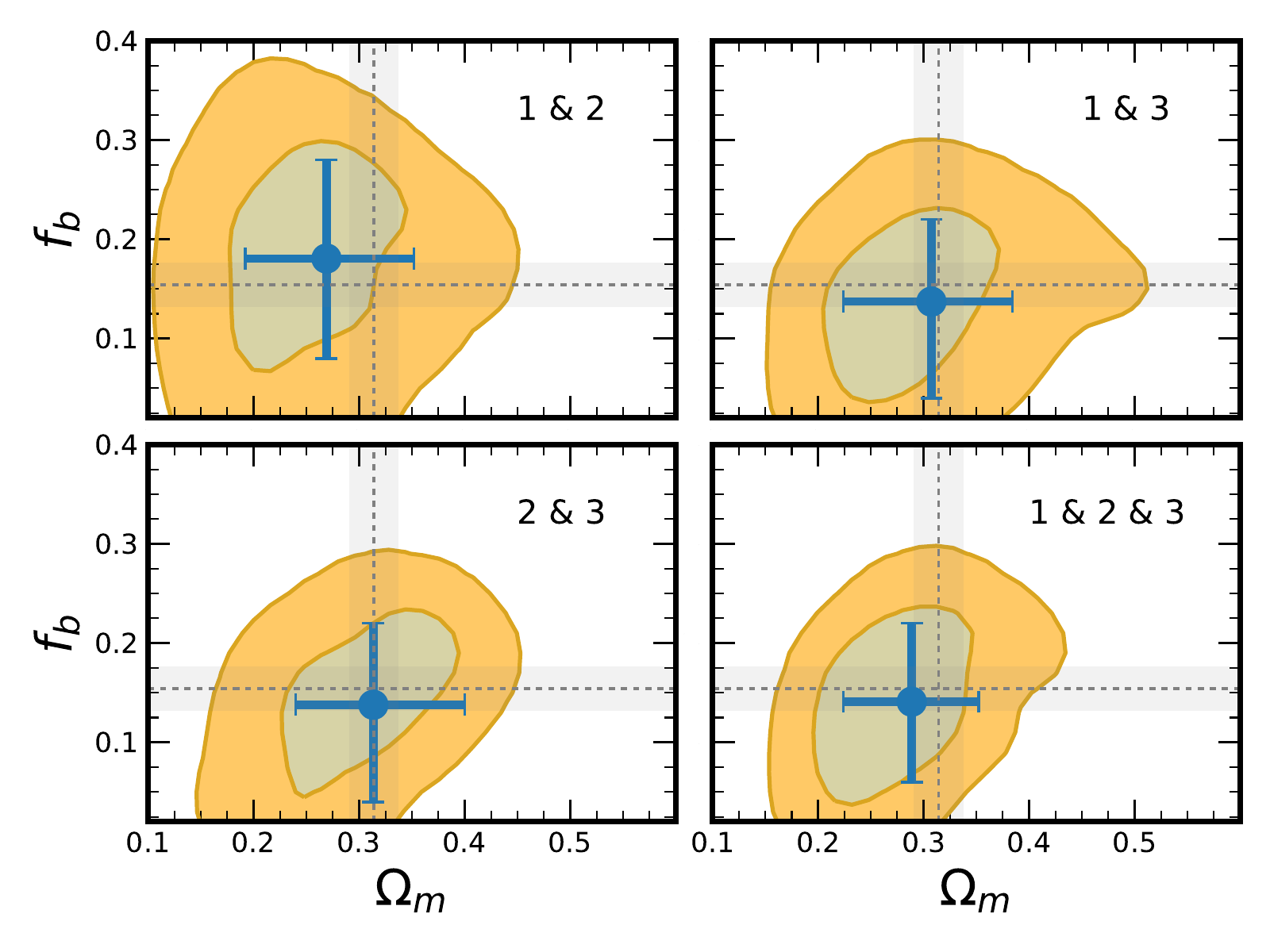}\\
  \includegraphics[width=8cm]{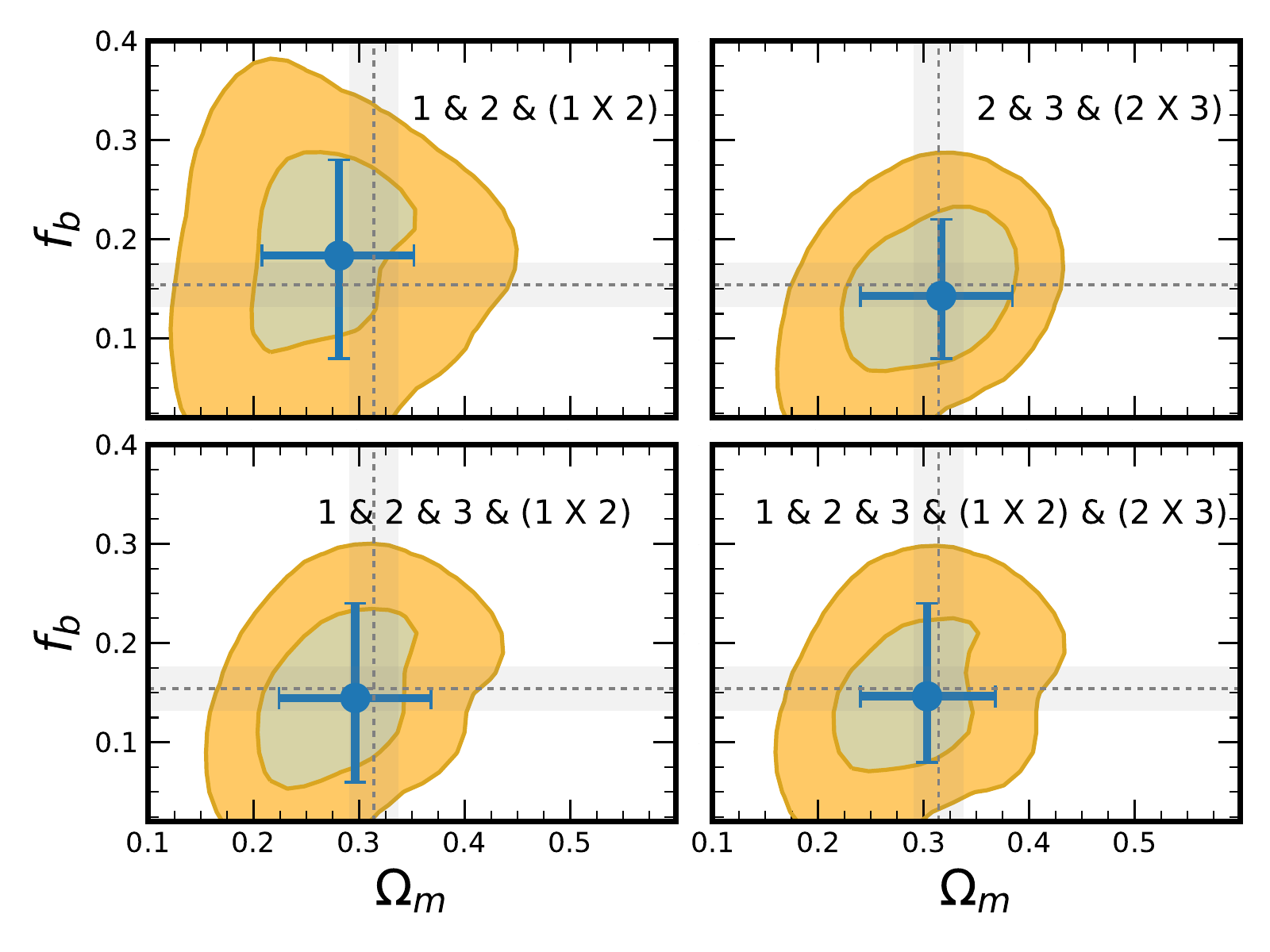}
  \caption{$68\%$ and $99\%$ confidence intervals  for $f_{b}$ and $\Omega_{\rm m}$ obtained after marginalizing over $\sigma_8$ and the bias parameters $b_i$.
  In the upper four panels we combine the auto-spectra from various redshift bins. The lower panels illustrate the effect of additionally using the cross-spectra for the constraints. Dotted vertical and horizontal lines show the Planck results and their 1-$\sigma$ errors (shaded region). 
   }\label{cl_cross}
\end{figure}

\begin{figure}
  \includegraphics[width=8cm]{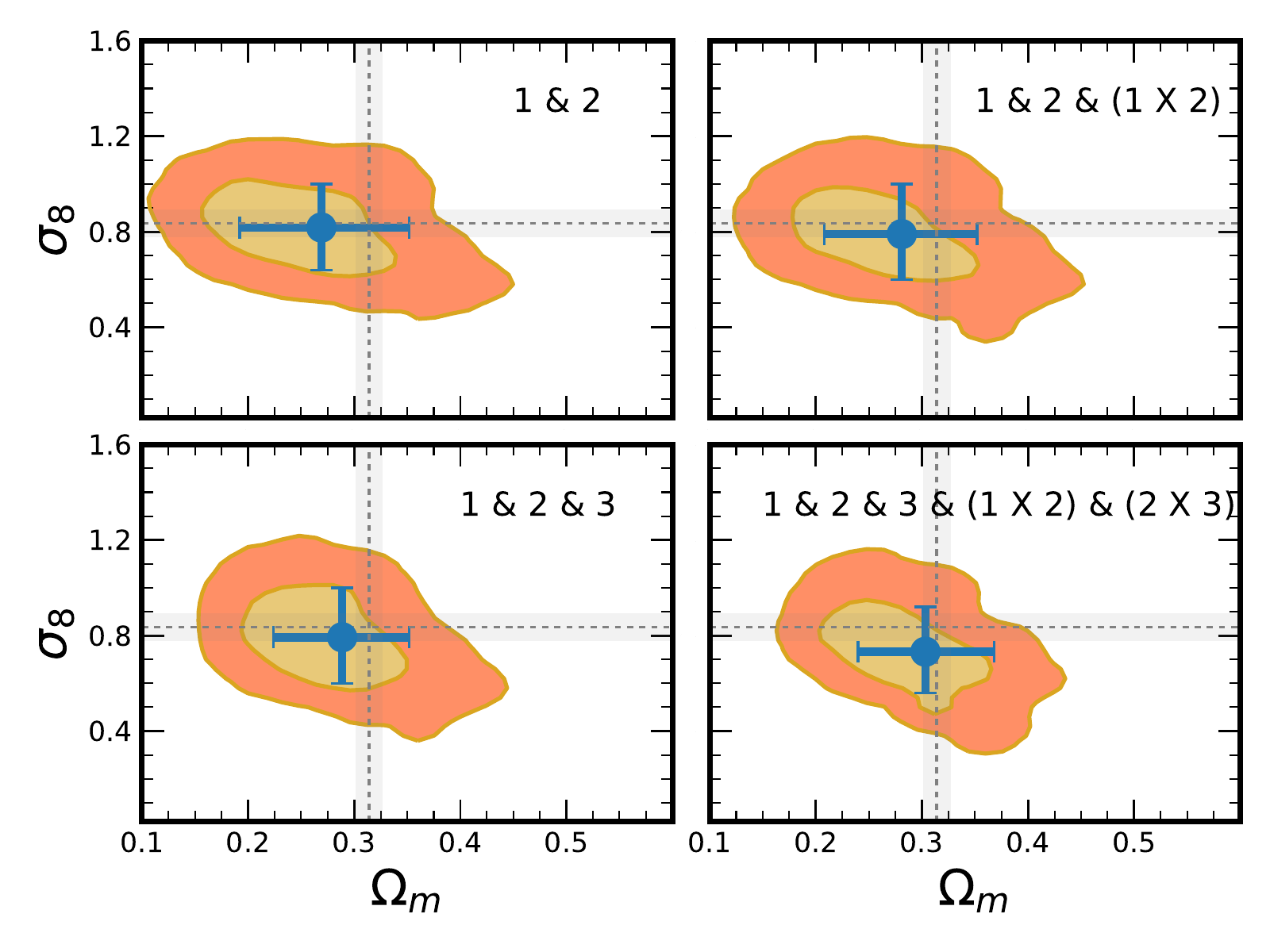}
  \caption{Similar as Fig.\ref{cl_cross} but for the set of parameters $\sigma_{8}$ and $\Omega_{\rm m}$.}\label{cl_crossb}
\end{figure}

\subsection{Multiple redshift bins}

In this Section we first combine the auto-correlation analyses performed in each bin to improve the constraints on the cosmological parameters. Then, we include the results obtained by cross-correlating 2MPZ galaxies in nearby bins, i.e. we also compute the angular cross-spectra between bins $1$ and $2$, and also $2$ and $3$. The cross-correlation between bins $1$ and $3$ is consistent with zero and will be ignored.

To combine these results we assume no correlation along the radial direction and test the goodness of this hypothesis {\it a posteriori}.
With this hypothesis we can compute the combined posterior probability  $\mathcal{P}_{ij}(\theta_{\rm cosmo}|\hat{C}_{ij})$ where $\theta_{\rm cosmo}=\{f_{b},\Omega_{\rm m}, \sigma_{8}\}$
in three steps: {\it 1)} We compute the posterior probability for each auto- or cross-angular spectra. {\it 2)} We marginalize each probability over the bias parameter (or bias parameters in case of cross-spectra) in the redshift bin. {\it 3)} We compute  $\mathcal{P}_{ij}(\theta_{\rm cosmo}|\hat{C}_{ij})$ by multiplying the various posterior probabilities together. 
The results are summarized in Fig.~\ref{cl_cross}, where we show the confidence levels in the $\{f_{b},\Omega_{\rm m}\}$ plane, analogous to those plotted in 
the upper panels of Fig.~\ref{cl_cosmo_2mpz}. To clarify the notation $i$ \& $j$ indicates that we combine information from auto-spectra in redshift bins 
$i$ and $j$, whereas $i \times j$ indicates that the cross-spectra between bins $i$ and $j$ have been included in the analysis. 
The upper four panels consider auto-spectra only and, among them, the bottom-right panel uses information from all the three redshift bins.
The four bottom panels are analogous to the upper ones except that they include cross-spectrum information. The values of the best fit parameters and their uncertainties are summarized in Table \ref{cvalues}.

Combining information from the different redshift bins does have an impact on the analysis. The errors on the estimated $\Omega_{\rm m}$ are reduced by a factor of 
about two. The largest improvement is obtained when the auto- and cross-spectra of 2MPZ galaxies in the outer redshift bins are included in the analysis.
A similar, significant improvement has also been found by \citet{2011MNRAS.412.1669T}. 
By comparison, the improvement on the baryon fraction error is less spectacular. Error bars are reduced by $10$-$30$ \% (again, the largest improvement 
is obtained using galaxies in the outer redshift bin) with no much benefit obtained by including cross-spectrum measurements.

By analogy with Fig.~\ref{cl_cross}, in Fig.~\ref{cl_crossb} we show the confidence contours in the $\{\sigma_8 \,, \Omega_{\rm m}\}$ plane, this time for auto-spectra only.
The fact that we obtain a constraint on $\sigma_{8}$ may seem in contradiction with the fact that, in the linear regime and 
with no RSD information from low multipoles (which, as emphasised earlier, we do not use), this parameter is fully degenerate with the linear bias parameters.
In fact, as anticipated, this degeneracy is broken by the fact that we use \texttt{Halo-fit} to model the APS
and that nonlinear contributions are not negligible at $\ell_{\rm MAX}=70$, especially in the first redshift bin. Not surprisingly, these constraints are not competitive with those obtained by CMB, 3D clustering analyses and cluster counts. We rather consider this measurement as a sanity check showing that the values of $\sigma_{8}$ obtained from our analysis (e.g. $\sigma_{8}=0.79^{+0.25}_{-0.19}$ from the combined analysis) are consistent with that obtained from Planck (grey strips) that we have used to infer the 2MPZ galaxy bias values.

We note that in the current implementation of \texttt{CLASSgal}, cross-power spectrum can be computed by modeling the galaxy $\dd N/\dd z_{\rm s}$ with either a Gaussian or a top hat function. We chose the first option despite the fact that, as can be deduced from Fig.~\ref{lf}, it does not provide a good fit to the galaxy redshift distribution, but it is certainly closer to reality than the top-hat option. We show in Appendix \ref{ap:dndz} that this choice does not introduce significant systematic errors.


\section{Discussion and Conclusions}\label{diss}

In this work we have performed a tomographic analysis in the spherical harmonic space to investigate the clustering properties of galaxies in the local Universe using the 2MPZ catalogue \citep{bilicki14}. Tomographic analyses have emerged as a complementary tool to investigate the LSS of the Universe when photometric, rather than spectroscopic, redshifts are available and a full study of the three dimensional distribution of objects is not possible.
Despite the fact that a significant amount of information is lost along the radial direction because of considerable \phz\ errors as compared to spectroscopy, the number of objects in photometric surveys is 
significantly larger than in spectroscopic ones. The former thus offer the possibility of densely sampling the LSS of the Universe over very large volumes which will not be easily available for the latter.

Several studies have explored the potential of the tomographic technique, its pros and cons, and demonstrated that it can already be applied to existing datasets to constraint cosmological parameters. While these constraints are not tight, they have the advantage of being complementary to those obtained from spectroscopic samples \citep{Percival2001,Cole2005,2009MNRAS.400.1643S,2011ApJ...736...59Z,Beutler2012,Ross2015,Howlett2015}. 
As a result, the tomographic technique is now regarded as one of the most promising tools to apply to next generation photometric redshift surveys like Euclid \cite[][]{laureijs} and LSST \citep[][]{LSST}
and new strategies are being proposed on how to combine information from spectroscopic and photometric samples (see e.g. \citealt{2017MNRAS.472L..40P} for a recent example).

We have used the tomographic technique to analyze galaxy clustering in the local Universe, bridging the gap between 2D clustering studies of large and wide photometric-only catalogs, such as 2MASS, and 3D clustering analyses performed with smaller and sparser spectroscopic samples, such as PSCz, 2MRS and 6dFGS. We are aware that this application stretches the method to its limits, since the combination of nonlinear effects, limited volume, uneven sky coverage, and other related issues severely limits the power of the method. Nevertheless, we decided to proceed because of the availability of the new, wide 2MPZ galaxy \phz\ dataset built upon the 2MASS photometric survey \citep[][]{bilicki14}.
Wide coverage is of paramount importance in local studies to maximize the volume of the survey and mitigate the impact of  the unavoidable cosmic variance. Good 
\phz\ calibration and small random errors are also highly desirable to efficiently slice up the volume in independent redshift shells.
2MPZ satisfies both these requirements since it allowed us to sample about $2.8\, \pi$ steradians, covering both the northern and southern hemispheres, with $\sim 700,000$ galaxies divided in three 
equal sized narrow redshift bins of width $\Delta z = 0.08$.

The results of our analysis can be summarized as follows:

\begin{itemize}

\item 3D clustering analyses have already been carried out in spectroscopic samples (2dFGRS, 6dFGS and SDSS) that partially overlap with 2MPZ.
With these results available, the first goal of the tomographic analysis is to provide a clustering-based, independent validation of the 2MPZ catalogue itself. The presence of anomalous features in the clustering statistics (APS in this case) would indicate potential issues in e.g. the survey photometry, redshift calibration etc, that should be further investigated.

The imprint of these potential systematic errors is expected to display a characteristic north-south pattern, both in Equatorial and in Galactic coordinates.
We extensively searched for smoking gun signatures by comparing results obtained independently in the various hemispheres and found no evidence of them in 
any of the statistics considered, namely the 1-point galaxy density probability distribution function, the APS and the cosmological parameters (baryon 
fraction, mass density, and galaxy {\it rms} number density fluctuations).

We checked that these tests are significant in the sense that the various hemispheres we have divided the 2MPZ into have similar areas and window functions, and therefore provide a similar amount of information.

We conclude that 2MPZ is suitable for clustering analysis.

\item We also looked for anomalous clustering power at $\ell<30$ to investigate the reality of the corresponding feature detected in the 2D clustering analysis of 2MASS galaxies brighter than $K=12.5$ by \citet[][]{2005MNRAS.361..701F}. The authors of that analysis suggest that such excess power and the presence of a large ``local hole'' fit in the same picture of a potential failure of the $\Lambda$CDM model. Our tomographic analysis does not support this claim, even though we find more power on large scales than predicted by a simple power-law APS model. This feature is more evident in the first redshift bin and at $\ell<15$, only partially overlapping with the range of $5<\ell<30$ where excess power was seen by \citealt{2005MNRAS.361..701F}, and it is robust to the flux cut. However, we find no tension between our results and the $\Lambda$CDM model, which instead provides a good match to our measured APS down to the largest angular scales probed by our analysis.

\item Performing a tomographic analysis in the local Universe has its own peculiarities. It should be designed as a balance between the need to maximize the cosmological information and that to reduce the systematic errors. 
The natural two-point statistics for an almost full-sky survey is the APS, that we estimated with the methods introduced by \citet[][]{1973ApJ...185..413P}. 
Having very large and homogeneous sky coverage guarantees a favourable window function, with reduced spurious correlation among multipoles. In our analysis we used mock 2MPZ catalogues to carefully investigate the impact of the window function and our ability to model 
its convolution effect on the underlying APS. 
The main effect of the mask is to remove power on large angular scales. The amplitude of the effect ranges between $5-10 \%$ for $\ell <10$.
We also showed that in our analysis we can account for this effect with better than $\sim 1$ \% accuracy. 
Nevertheless, and taking into consideration the large cosmic variance at low multipoles, we decided to adopt a conservative approach 
and focus our analysis on the multipoles $\ell \geq 10$. We verified that pushing our analysis down to the first $\ell$-bin does neither significantly modify nor reduce the statistical errors in our results.
Instead, our results suggest that including small multipoles can generate systematic errors.

Nonlinear effects, both in galaxy bias and underlying dynamics, are also important in the local Universe and may have an impact 
on fairly large angular scales. They are also difficult to model accurately.
Instead of attempting to model these effects  {\it a priori}, we assessed their impact {\it a posteriori}. Guided by 
 \texttt{Halo-fit} \citep[][]{2003MNRAS.341.1311S, 2012ApJ...761..152T}, which provides an indication on the scale of nonlinearity, we simply verified the robustness of our results to
 the choice of the maximum multipole $\ell_{\rm MAX}$ to which we extend our analysis. As a result we decided to adopt a value $\ell_{\rm MAX}=70$ and found that 
 we can safely push our analysis  to $\ell_{\rm MAX}=100$ except for the lowest redshift bin, for which we found a hint of systematic effects if such small scales ($\ell>70$) were included, and the highest redshift bin, which for scales $\ell\gtrsim  70$ is shot-noise dominated.
 
Finally, in the tomographic analysis one needs to account for the impact of random \phz\ errors that displace objects along the line of sight. These displacements mean that a galaxy sample selected in a \phz\ bin is contaminated by objects at higher or smaller redshifts.
The narrower the bin, the larger the contamination.
A tradeoff needs to be found between minimizing the contamination level and maximizing the number of bins to take the full advantage of the tomographic approach \citep[e.g.][]{BlakeBridle2005,2012MNRAS.427.1891A}.
We have investigated this issue with the help of the 2MPZ mock galaxy catalogues and found that
considering objects in the redshift range $z=[0,0.24]$ and dividing the sample into three equally spaced bins represents good compromise. The residual contamination effect is accounted for in the likelihood analysis using different approaches that, as we have verified, provide
very similar results.

\item To estimate the statistical errors and their covariance we have created 1000 catalogues of mock 2MPZ galaxies with a lognormal density distribution function,
 \texttt{Halo-fit} angular power spectrum of a $\Lambda$CDM model, Gaussian \phz\ errors, and the same geometry as the real survey.
 The angular power spectra measured in each of the $1000$ mocks for each redshift bin were used to compute the covariance matrices of the angular auto- and cross-power spectra. 
 This is a rigorous but computationally intensive approach that, for the sake of accuracy, should be repeated for any cosmological model considered in the likelihood analysis.
 To check whether other, less time-consuming approaches could be adopted without compromising the quality of the results, we have computed errors with two alternative methods: a jackknife resampling technique and the analytic Gaussian assumption.
 In our analysis we compared the errors and addressed the robustness of the likelihood analysis to the type of error estimate.
 We found that the three methods provide very similar error estimates. The exception is the jackknife technique, which systematically overestimates the uncertainties, by $\sim 20$\%, although
in the first redshift bin only.

As a result we decided to use Gaussian errors, similarly as in the previous tomographic analyses of SDSS samples by \cite{2007MNRAS.374.1527B} or \cite{2011MNRAS.412.1669T}.

 \item We have used the public code \texttt{MontePython} to Monte Carlo sample the posterior probability of selected cosmological parameters, namely the baryon fraction, the mean mass density
and the combination of galaxy bias and {\it rms} mass density fluctuation, given the estimated angular auto- and cross-spectra in the three redshift bins.
Flat priors were set on the dark matter density, baryon density, primordial spectral amplitude and effective linear galaxy bias at the mean redshifts of the three bins.
All remaining cosmological parameters were fixed at their Planck values \citep[][]{2014A&A...571A..16P}. 

From the analysis of the auto-spectra in each redshift bin independently, we measured $f_{\rm b}$ and  $\Omega_{\rm m}$ and found  that they are in agreement with the reference $\Lambda$CDM model. 
However, uncertainties are large; 1-$\sigma$ errors on $\Omega_m$ are of the order of $50\%$, and even larger for the baryon fraction.

Combining different auto-spectra under the hypothesis of no radial correlation among the bins significantly improves the results and reduces the relative errors to $\sim 25\%$ for $\Omega_{\rm m}$
and to $\sim 50\%$ for $f_{\rm b}$. Additional information from the cross-spectra does not bring significant improvements (1-$\sigma$ errors on $\Omega_m$ drop to $20\%$), which indicates that cross-power 
is indeed small and the hypothesis of negligible radial correlation among the bins is indeed a reasonable one.

Our error bars are about twice as large as in the similar tomographic analysis of the SDSS samples such as \citet[][]{2011MNRAS.412.1669T}. This is not entirely unexpected: it reflects the large cosmic variance which is typical of cosmological investigations of the local Universe, further exacerbated by the limited multipole range accessible to our analysis.
 A denser sampling of a more linear density field over a significantly larger volume, as in the SDSS case, would significantly improve the quality of the analysis. 
 This is the key to the success of the tomographic analyses that will be performed on forthcoming datasets like the Euclid photometric catalogue \cite[][]{laureijs} and the LSST galaxy sample \cite[][]{LSST}. We note however that such studies could be also attempted with already existing deep wide-angle \phz\ datasets, such as WISE $\times$ SuperCOSMOS \citep{bilicki16} or SDSS DR12 \citep{2016MNRAS.460.1371B}.
 
Driven by the need to keep the number of free parameters small, we have restricted our analysis to the regime in which galaxy bias is close to the linear model. 
As a result, from the APS in each redshift bin we have constrained the combination $b_i\sigma_8$, which we used to estimate the effective bias parameters of 2MPZ galaxies after fixing $\sigma_8$
to its Planck value. We were able to estimate such effective bias parameters with fairly good precision ($15-20\%$) and found that $b_{\rm eff}(z)$ increases by $\sim 60\%$ from the first redshift bin of median \phz\ of $\langle z_{\rm p}\rangle =0.05$ to the third one with $\langle z_{\rm p}\rangle =0.19$.
This rapid change simply reflects the apparent magnitude-limited nature of the catalogue, which selects objects increasingly brighter intrinsically at larger redshifts.

Bias parameters can be marginalized over when combining auto- and cross-spectra measured in different redshift bins and thanks to the nonlinearities quantified by the 1-halo term within the \texttt{Halo-Fit} framework, which breaks the degeneracy between $b_i$ and $\sigma_8$. The resulting $\sigma_8$ value, though not at all competitive with those obtained with other probes, is nevertheless in agreement with the Planck value. This constitutes a useful sanity check for our analysis and justifies {\it a posteriori} our procedure to estimate the galaxy bias.

\end{itemize}

The 2MPZ APS contains not only the cosmological information we have described in this paper. In a forthcoming paper we will explore the astrophysical content in the clustering signal by interpreting our measurements in the context of the halo model \cite[e.g.][]{2000MNRAS.318..203S, 2002PhR...372....1C,2003ApJ...593....1B, 2004ApJ...609...35K, 2005ApJ...633..791Z} hence generalizing the results of \cite{2018MNRAS.473.4318A} obtained for much shallower ($\langle z \rangle = 0.03$) 2MASS Redshift Survey. We will combine the information from the APS with the 2MPZ luminosity function. We will also use the 2MPZ catalogue and the machinery developed in this paper to perform a detailed clustering-based cosmography analysis of the local Universe.

Finally, together with this paper we provide upon request a user-friendly version of our power spectrum estimation code \texttt{H-GAPS} (\textit{\texttt{Healpix}-based Galaxy Angular Power Spectrum})\footnote{\url{https://abalant.wixsite.com/abalan/to-share-1}}, which allows for the computation of the power spectrum and the mixing matrix for an input galaxy catalogue and a \texttt{Healpix} mask.

\section*{Acknowledgments}
We acknowledge Chris Blake for the careful reading of our manuscript and the valuable comments. ABA acknowledges financial support from the Spanish Ministry of Economy and Competitiveness (MINECO) under the Severo Ochoa program SEV-2015-0548. MB is supported by the Netherlands Organization for Scientific Research, NWO, through grant number 614.001.451, and by the Polish National Science Center under contract UMO$-2012/07/$D$/$ST$9/02785$. EB is supported by MUIR PRIN 2015 ``Cosmology and Fundamental Physics: illuminating the Dark Universe with Euclid'' and Agenzia Spaziale Italiana agreement ASI/INAF/I/023/12/0. ABA, MB and EB acknowledge support of the Ministry of Foreign Affairs and International Cooperation, Directorate General for the Country Promotion (Bilateral Grant Agreement ZA14GR02 - Mapping the Universe on the Pathway to SKA). This research has made use of data obtained from the SuperCOSMOS Science Archive, prepared and hosted by the Wide Field Astronomy Unit, Institute for Astronomy, University of Edinburgh, which is funded by the UK Science and Technology Facilities Council. This work has made use of \texttt{TOPCAT} \citep{2005ASPC..347...29T}.


\bibliographystyle{mnras}
\bibliography{refs}  


\appendix


\section{The estimator of the angular power spectrum}\label{sec:estimator_ap}
\begin{figure}
\vspace{-5cm}
  \includegraphics[width=9cm]{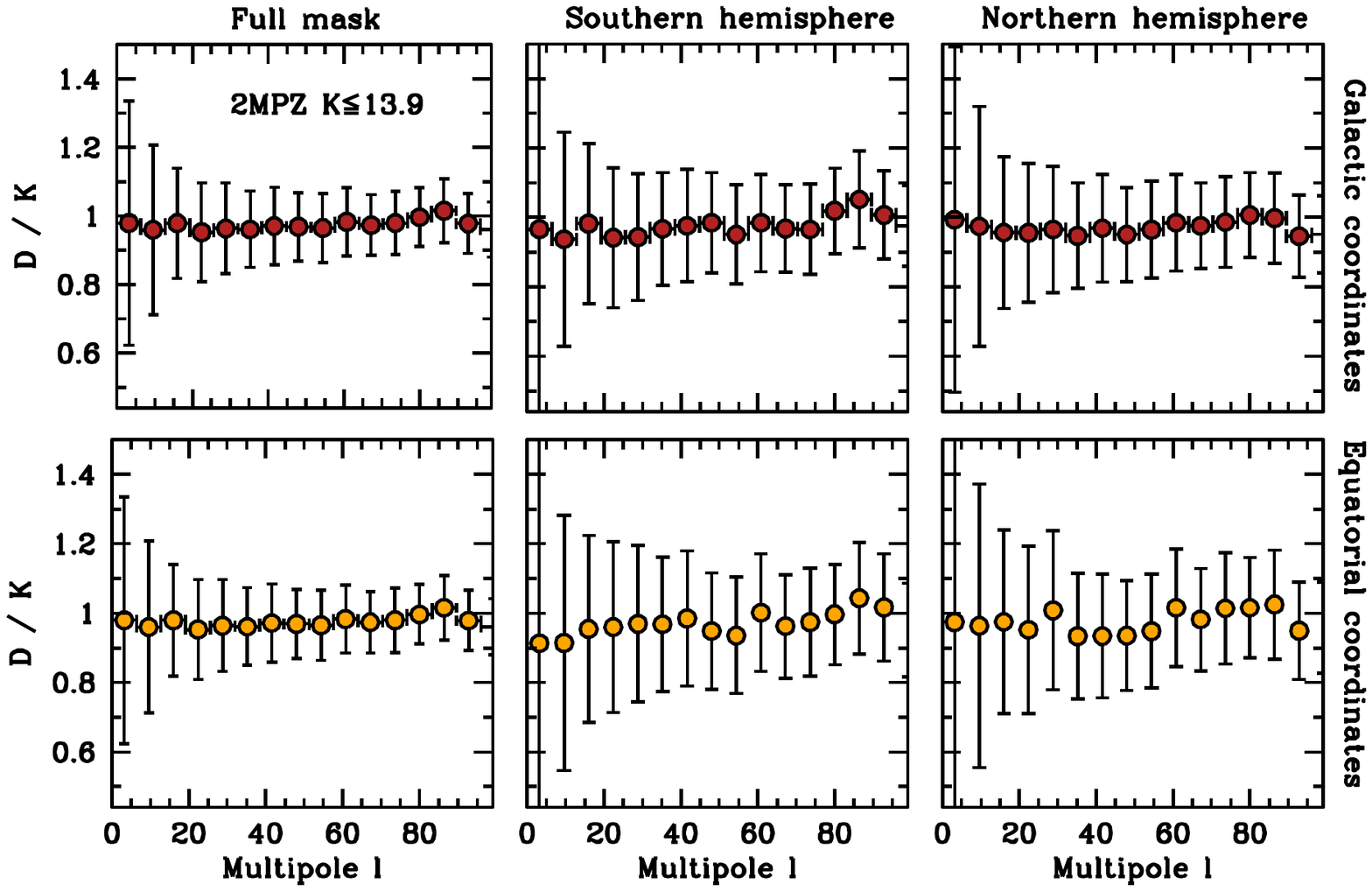}
\vspace{-2.5cm}
\caption{Ratio between the APS of the 2MPZ sample (full redshift range) measured with the estimator $\hat{D}$ defined by Eq.~(\ref{dl}) and that obtained with the estimator $\hat{K}$ from Eq.~(\ref{klt}), for different hemispheres and coordinate systems. The error bars represent the \textit{rms} scatter of the 2MPZ mock catalogues described in Sect.~\ref{smocks}.}\label{cl_DK}
\end{figure}

Together with the estimator $\hat{K}$ defined in Eq.(\ref{klt}), \citet[][]{1973ApJ...185..413P} also introduced an estimator for the APS of the form:
\be\label{dl}
\hat{D}^{ij}_{\ell}=\frac{1}{2\ell+1}\sum_{m=-\ell}^{m=+\ell} \frac{|a^{i}_{\ell m}a^{* j}_{\ell m}|}{J_{\ell m}}- \frac{1}{\bar{\sigma}_{i}}\delta^{K}_{ij},
\ee
where $a_{\ell m}$ represent the spherical harmonic coefficients defined in Eq.~(\ref{almd}, the second term is the shot-noise correction and 
\be\label{jlm}
J_{\ell m}\equiv \int  \,M(\hvr)|Y_{\ell m}(\hvr)|^{2}\dd\,\hvr, \ee
with $M(\hvr)$ the angular mask. As in Eq.(\ref{ucl}), the ensemble average of this estimator introduces the mixing matrix of the form
\be\label{rl2}
\tilde{R}_{\ell \ell'}=\frac{1}{2\ell+1}\sum_{m=-\ell}^{m=+\ell}\sum_{m'=-\ell'}^{m'=\ell'} J^{-1}_{\ell m} \left |\int Y^{*}_{\ell m}(\hvr)M(\hvr)Y_{\ell' m'}(\hvr)\dd \hvr \right |^{2},
\ee
which, unlike the mixing matrix of Eq.~(\ref{mm}), cannot be written in terms of $3j$ Wigner symbols. It is important therefore to use the mixing matrix appropriate to the estimator adopted. At the level of a likelihood analysis, an incorrect choice might lead to a systematic effect in the constraints of cosmological (or astrophysical) parameters. Such systematic is clearly reduced as long as the sample covers larger fractions of the sky, in which case the measurements obtained with the estimators $\hat{D}$ of Eq.~(\ref{dl}) and $\hat{K}$ come to closer agreement. Given the sky fraction covered by the 2MPZ galaxy catalogue, the difference between these two estimators are below the error bars assigned to the measurements, as is shown in Fig.~\ref{cl_DK}.

\section{The impact of the 2MPZ mixing matrix and pixelization} \label{sec:mask}
\begin{figure}
\vspace{-2cm}
  \includegraphics[width=8.5cm]{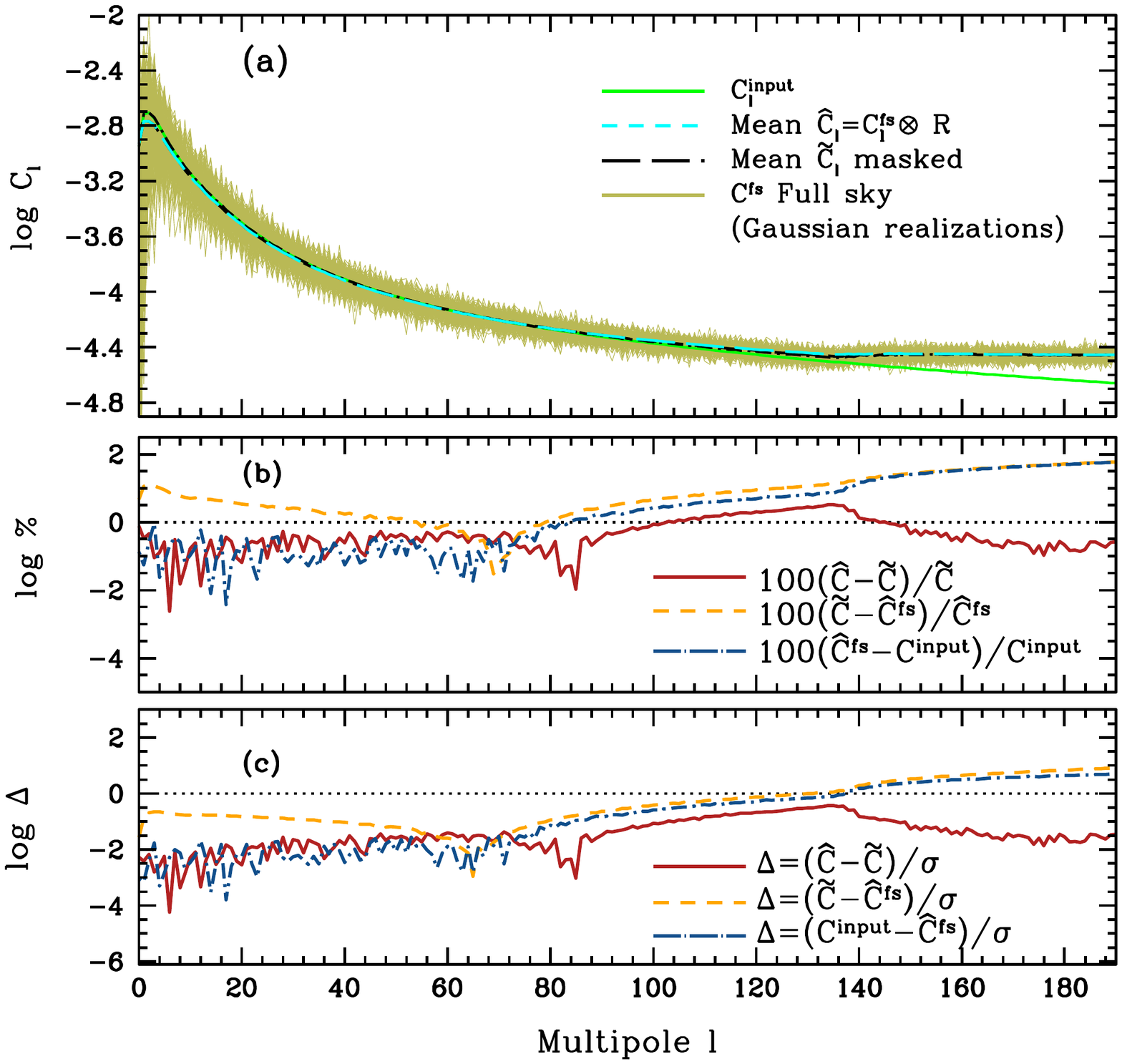}
\vspace{-2cm}
  \caption{Panel (a) shows the input power spectrum $C^{\rm input}_{\ell}$ (solid red line) and the $10^{4}$ Gaussian realizations (overlapping light green curves). The short-dashed blue line illustrates the mean of the convolution of the $10^{4}$ full-sky APS with the 2MPZ mixing matrix ($\hat{C}_{\ell}$). The long-dashed orange line presents the APS from the $10^{4}$ maps upon which the 2MPZ mask has been imposed ($\tilde{C}_{\ell}$).  Panel (b) shows the percentage difference between these spectra (exactness), and panel (c) provides the difference among these spectra in units of the statistical error. }\label{cl_mask}
\end{figure}

The measured 2MPZ APS is different from the true one because of a number of effects. Here we explore those introduced by the survey geometry and by the map resolution.
To assess their impact we adopt the following procedure.

\begin{itemize}
\item Given a theoretical APS, $C^{\rm input}_{\ell}$, we generate a set of $10^{4}$ Gaussian distributed harmonic coefficients $a_{\ell m}$, with zero mean and variance $(C^{\rm input}_{\ell})^{1/2}$.
\item For each realization, a full-sky overdensity map $\delta_{\rm fs}(\hvr)$ is created using the \texttt{alm2map} routines in \texttt{Healpix}. We measure the power spectrum for each of these $10^{4}$ full-sky maps and estimate its mean ($\hat{C}_{\ell}^{\rm fs}$) and variance.
\item In parallel, for each realization we use Eq.~(\ref{cij2}) to compute the the convolved APS, $\hat{C}_{\ell}$, using the mixing matrix described in Sect.~\ref{sec:mix}.  
\item We apply the geometry mask $M(\hvr)$ to the full-sky map to obtain the masked overdensity field and estimate its APS $\tilde{C}_{\ell}$. 
\end{itemize}
  
Panel (a) in Fig.~\ref{cl_mask} shows the different power spectra obtained with this procedure. Panel (b) shows the relative differences between the three APS. The most relevant is the red solid curve that compares the masked and the convolved spectrum. Panel (c) shows these differences in units of statistical errors, $\sigma$, estimated from the scatter among the mocks.
From these comparisons we conclude that:

\begin{itemize}
\item  The effect of the 2MPZ mixing matrix, quantified by the difference between $C^{\rm input}_{\ell}$ and $\tilde{C}_{\ell}$ (dashed curves), is significant on large scales. Its amplitude of $10\%$ at $\ell=2$ decreases with $\ell$ and drops to $1\%$ at $\ell=50$. This systematic effects is however small, less than $10\%$, compared to the statistical error.

\item The difference between the masked power $\tilde{C}_{\ell}$ and the convolved $\hat{C}_{\ell}$ (solid line in Fig.~\ref{cl_mask} panel b) is $\leq 1\%$ and much smaller than the statistical errors. We conclude that the estimated mixing matrix and its convolution with the true power spectrum do match the APS measured from the mock 2MPZ map.

\item  The comparison between the input power spectrum $C^{\rm input}_{\ell}$ and the measured full-sky spectrum $\tilde{C}^{\rm fs}_{\ell}$ quantifies the impact of the map resolution. As expected the effect is significant on the angular scales of the pixel (i.e. $\ell \gtrsim 120$). Its amplitude of $\sim 1\%$ at $\ell \sim 80$ increases with $\ell$ and matches the statistical error at $\ell\sim 140$.

\end{itemize}
  These results can be used to set the multipole range in which to compare the model and measured APS. To be conservative, we discard the multipoles below $\ell=10$ (i.e. we discard our first two $\ell$-bins), where the impact of the mask is significant, and multipoles above $\ell=100$, to avoid map resolution effects. Other effects like shot-noise and nonlinearity will further decrease this upper limit.


\section{Robustness tests for the likelihood analysis }\label{check}

\begin{figure}
  \vspace{-0.5cm}
  \includegraphics[width=8cm]{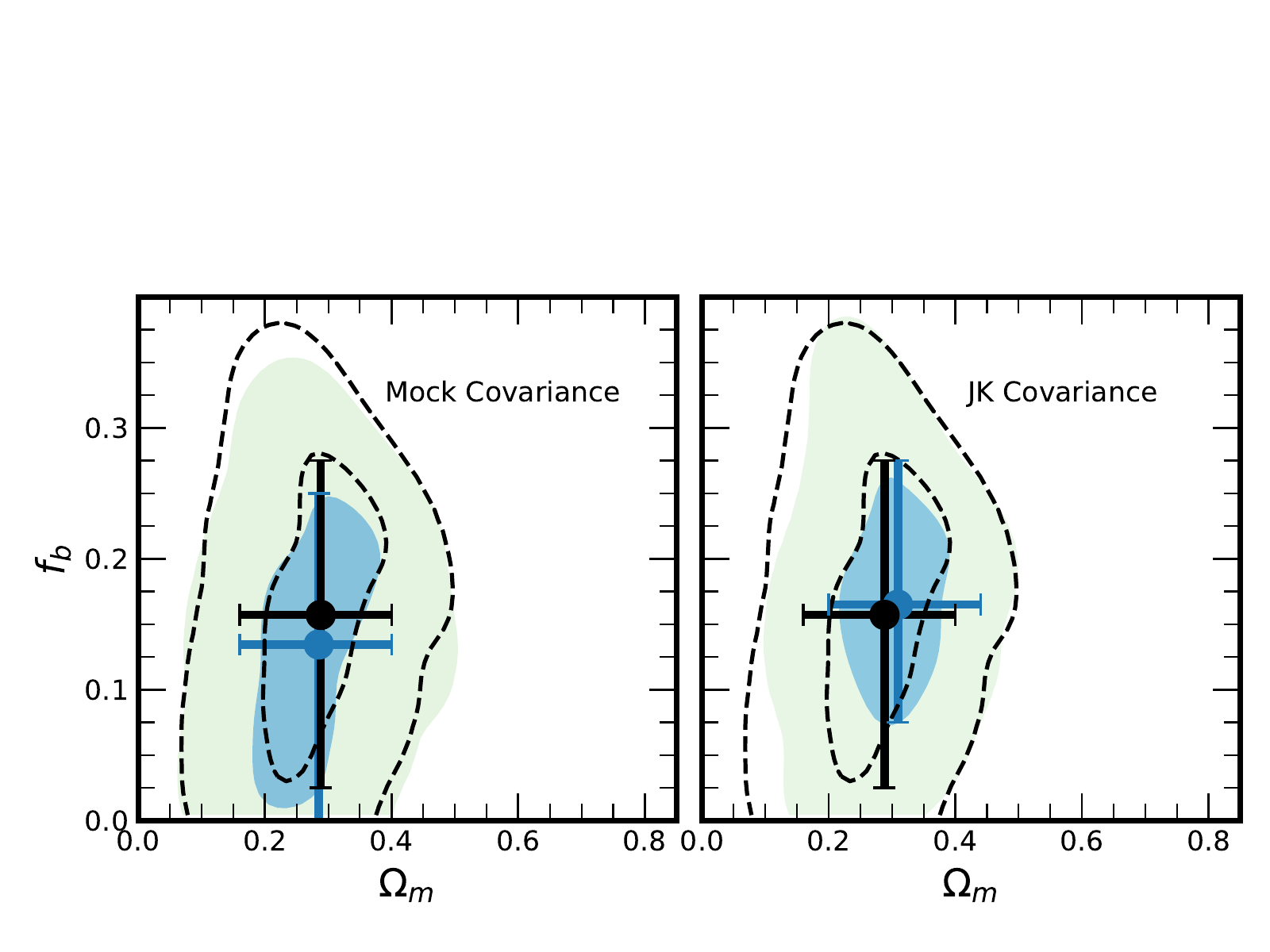}
  \caption{Sensitivity to error estimates. 
   Best fit values with marginalized error bars and $68\%$ and $99\%$ confidence regions for $f_{b}$-$\Omega_{\rm m}$. Dashed contours: Gaussian errors. 
   Continuous contours: covariant errors from lognormal mocks (left) and jackknife errors (right). Black dots with error bars: Gaussian case. Blue dots with error bars: 
  covariance matrix (left) and jackknife errors (right). All results refer to the second 2MPZ redshift bin.
   }\label{cl_tests}
\end{figure}

In this Section we check the sensitivity of the results to the input of the likelihood analysis, namely the covariant errors in the binned spectra and the galaxy redshift distribution used to model the angular spectra. We also check the robustness to splitting the samples into two hemispheres. Instead, the sensitivity to the minimum and maximum multipoles used in the analysis is discussed in the main text.

\begin{figure}
\center
  \includegraphics[width=6cm]{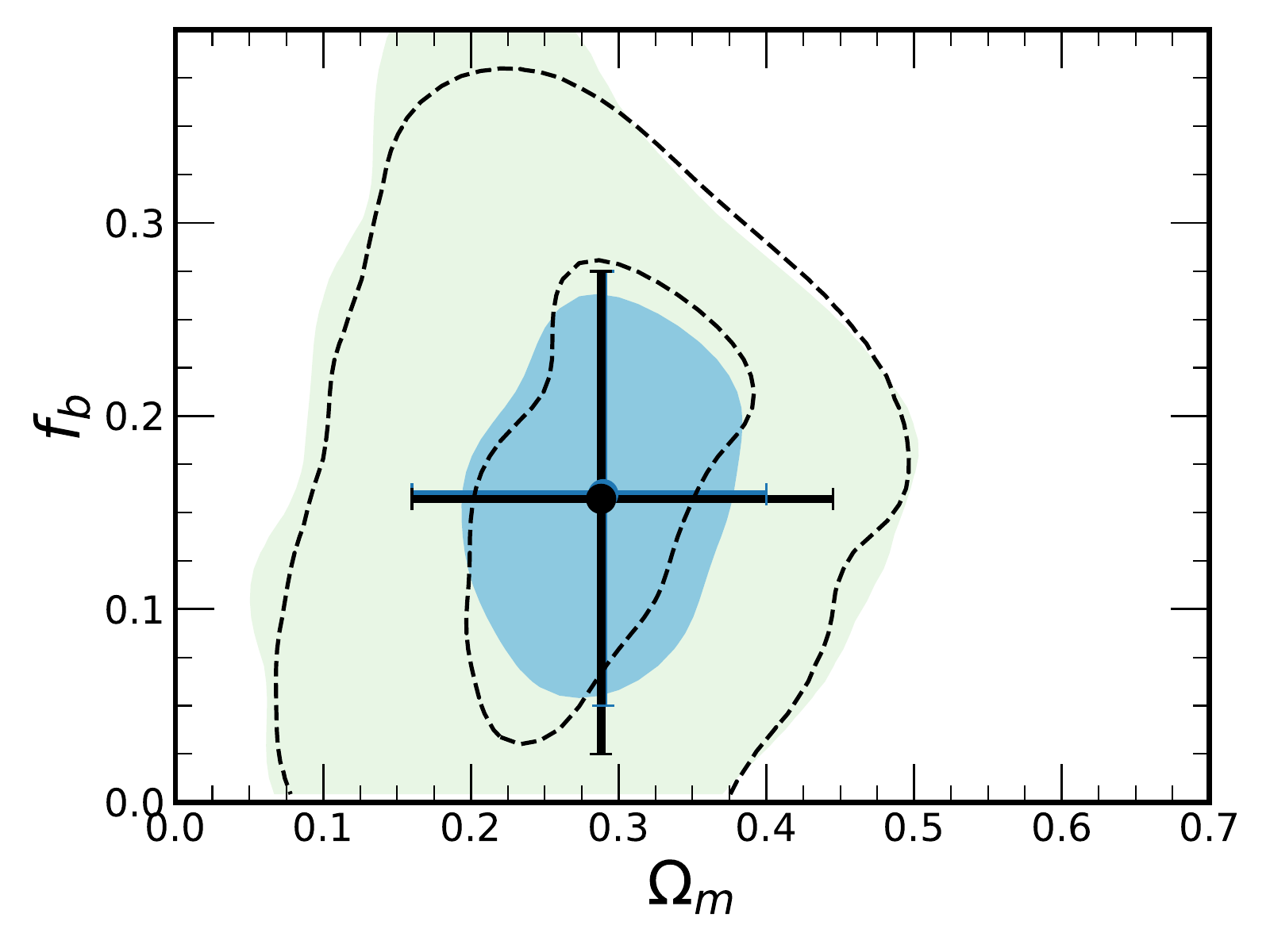}
  \caption{Sensitivity to galaxy redshift distribution.
   Best fit values with marginalized error bars and $68\%$ and $99\%$ confidence regions for $f_{b}$-$\Omega_{\rm m}$. Dashed contours and black dots with error bars: true $\dd N/\dd z$ from 
   convolution.  Continuous contours and blue dots with error bars: Gaussian $\dd N/ \dd z$ case. 
       All results refer to the second 2MPZ redshift shell.}\label{cl_testz}
\end{figure}

\subsection{Sensitivity to the estimated errors}\label{ap:err}
In our analysis we have used three different methods to estimate the error of the 2MPZ power spectrum and their covariance: analytic Gaussian errors, jackknife procedure,
and covariance matrix from the lognormal mock catalogues. As we have discussed in Sec.~\ref{sec:errcomp}, we decided to adopt Gaussian errors having verified that the 
results do not change significantly when adopting any of the two other methods. Here we show that the estimated cosmological parameters are robust to the type of error considered. 

Figure \ref{cl_tests} shows the confidence contours in the $f_{b}$-$\Omega_{\rm m}$ plane together with their best fit values (dots) and the marginalized 1-$\sigma$ error bars.
We only show the results obtained in the second redshift bin since they are representative for the other two bins.
The black dot and dashed contours refer to the baseline model of Gaussian errors.
In the left panel the blue dot and the filled  contours show the results obtained when the likelihood is computed using the full covariance matrix from the mocks.
They are remarkably similar to the baseline case, showing that ignoring covariance does not introduce any appreciable difference, apart from slightly reducing the 
size of the errors.
The same considerations apply to the jackknife errors (panel to the right).

We conclude that our choice to adopt Gaussian errors is entirely justified and does not introduce significant systematic effects.

\subsection{Sensitivity to the galaxy redshift distribution}\label{ap:dndz}

To model the APS in the generic redshift bin one needs to specify the true (i.e. spectroscopic) galaxy redshift distribution in that bin. In Sect.~\ref{sec:photoz} we described the procedure to infer the true redshift distribution of 2MPZ galaxies in a \phz\ bin with sharp boundaries.
These distributions are shown in Fig.~\ref{lf} and are characterized by a significant skewness and kurtosis.
In this Section we want to check what is the impact of using a Gaussian model to describe those distributions.
The rationale behind this test is that when considering joint likelihood involving cross-power spectra, the current implementation of \texttt{CLASSGal} only accepts 
top-hat or Gaussian redshift distributions in the different redshift bins. Hence, in order to compute posterior distributions (as in Fig.~\ref{cl_cross}), only the auto-power spectra are computed using the results from Sect.~\ref{sec:photoz}, while the cross-power spectra are derived using a Gaussian redshift distribution.

Figure \ref{cl_testz} shows that despite providing a poor fit to the actual redshift distribution, adopting a Gaussian model has 
very little impact on the final results. This is illustrated for the second \phz\ bin, but the same results are found also in the first bin. We then extrapolate this result and conclude that the modeling of the cross-power spectra between the first and second redshift bins based on Gaussian fits for the redshift distribution does not introduce any significant systematic effect. For the third bin, a $\lesssim 1-\sigma$ systematic deviation in the measurement of $\Omega_{\rm m}$ appears when using the Gaussian fit in the auto-power spectrum analysis. Given that we do not use the cross-correlation between bins $1$ and $3$ (which is compatible with zero), the only potential systematic effect affecting the results from Sect.~\ref{sec:lik} is in an incorrect model of the cross-power between bins $2$ and $3$. Nevertheless, by inspection of Fig.~\ref{cl_cross} we see that adding the information from cross-power spectrum between bins $2$ and $3$ does not introduce significant systematic errors in the measurements of cosmological parameters.

\begin{figure}
  \includegraphics[width=8cm]{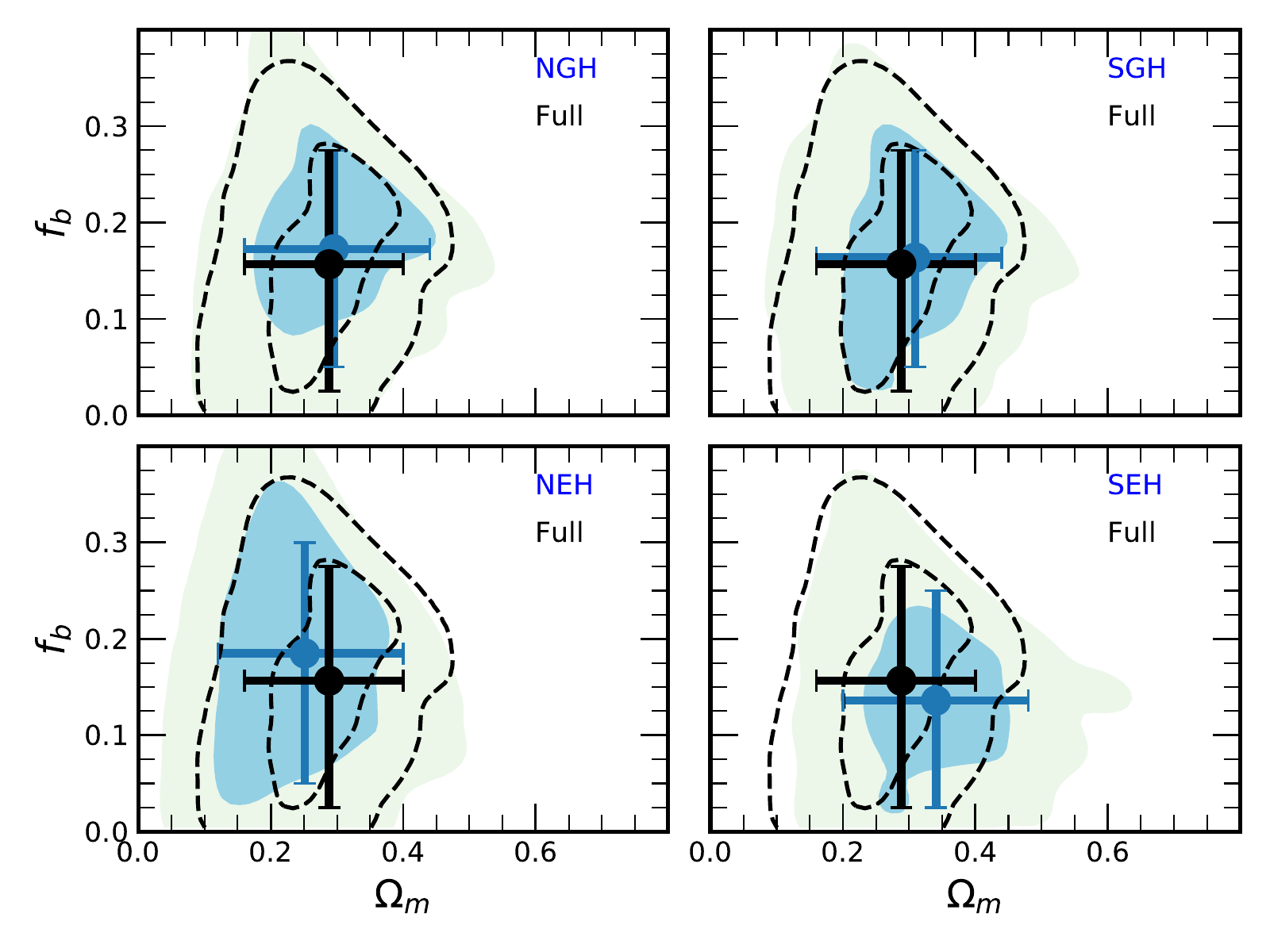}
  \caption{The $f_{b}$-$\Omega_{\rm m}$ parameters obtained in four hemispheres: North Galactic (top left), South Galactic (top right),
North Equatorial (bottom left), South Equatorial (bottom right). The dashed curves show the reference case of the all-sky 2MPZ sample. All contours are computed in the second redshift bin.}\label{cl_hemis}
\end{figure}

\subsection{Sensitivity to the split between North and South hemispheres}
In Sect.~\ref{sec:description} we discussed that 2MPZ is potentially prone to north-south systematic effects both in Galactic and Equatorial coordinates.
In the main text we searched for such effects in the 1-point overdensity PDF. Here we extend that search and look for systematic differences in the estimated cosmological parameters.
The results are shown in Fig.~\ref{cl_hemis}. We find no significant differences between the $\{f_{b}$, $\Omega_{\rm m}\}$  values estimated in the full sample and these obtained from the four hemispheres.

\end{document}